# Quasi-diagonal Inhomogeneous Closure for Classical and Quantum Statistical Dynamics

**Jorgen S. Frederiksen**[1,2]

[1]CSIRO Oceans and Atmosphere, Aspendale, Victoria, 3195, Australia;
[2]Monash University, Clayton, Victoria 3800, Australia.
E-Mail: Jorgen.Frederiksen@csiro.au

**Abstract:**

The Quasi-diagonal Direct Interaction Approximation (QDIA) closure equations are formulated for inhomogeneous classical and quantum fields interacting through dynamical equations with quadratic nonlinearity and with first or second order time derivatives. Associated more complex inhomogeneous DIA and Self-Energy closure equations are expounded as part of the derivation. The QDIA employs a bare vertex approximation and is only a few times more computationally intensive than the homogeneous DIA. Examples of applications to turbulent classical geophysical and Navier Stokes fluids, including non-Gaussian noise, to classical and quantum Klein Gordon equations with $g\phi^3$ Lagrangian interaction, and to coupled field-auxiliary field equations associated with $\lambda\phi^4$ Lagrangian interaction, are presented.

**Keywords:** inhomogeneous closures; classical fields; quantum fields; non-equilibrium statistical dynamics

## 1. Introduction

Statistical dynamical closure theories for classical and quantum fields were initially developed somewhat independently. Very elegant and general formalisms for quantum fields – particularly quantum electrodynamics – were formulated through the functional equations of Tomonaga (1946) and Schwinger (1948a, b, 1951a, b, c, 1953) and the equivalent diagrammatic techniques of Feynman (1949); see also Dyson (1949). In classical statistical fluid dynamics, the development of the direct interaction approximation (DIA) for homogeneous turbulence by Kraichnan (1959) was a pioneering advance. The DIA is a bare vertex approximation (BVA) for non-equilibrium and time-dependent fluids while the quantum field theories included more general interactions but were concerned primarily with scattering and with equilibrium and in-out states. Wyld (1961) and Lee (1965) used a diagrammatic approach that led to statistical dynamical equations for classical fluid dynamics and magneto-hydrodynamics to fourth and sixth order in perturbation theory, respectively, and thus included vertex corrections. Martin, Siggia and Rose (1973; hereafter MSR) generalized Schwinger's functional formalism to time-dependent classical fields. They recognized that the formalism needed to include the adjoint equation for the classical field in order to represent the response function that in general is not related to the two-time covariance through the fluctuation-dissipation theorem. MSR argued that their formalism produced additional vertex functions to those of Wyld (1961). However, recently Berera et al. (2013) have found, on correcting some minor errors, in both the diagrammatic and functional formalisms, that they indeed agree to fourth order. Rose (1974) generalized the MSR formalism to include additive random forcing and showed that non-Gaussian noise and non-Gaussian initial conditions could be included with additional interactions through so-called 'spurious' vertices. The Feynman path integral formulation (Feynman and Hibbs 1965) was subsequently used to further generalize the MSR formalism for classical systems to more complex interactions and additive and multiplicative random forces by Phythian (1977) and Jensen (1981); see also Krommes (2002).

In the early days of strong interaction hadron particle physics there was a focus on obtaining closure for the scattering amplitude between in and out states through dispersion relations such as the Mandelstam (1958, 1959) double spectral representation (Barut 1967; Frederiksen 1975; Frederiksen et al. 1975; Atkinson et al. 1976). The analyticity properties of scattering amplitudes were determined through Cutkosky (1960) rules and consistency of the Mandelstam representation with Feynman graphs within $g\phi^3$ theory was also established directly (Frederiksen and Woolcock 1973a, b; Frederiksen 1974a, b). These approaches are still employed such as in the recent study of light-by-light hadronic scattering by Colangelo et al. (2015).

In the case of classical high Reynolds number strongly turbulent flows the focus has, until recently, been on homogeneous, and generally isotropic, turbulence (McComb 2014) which is a somewhat idealized problem but it does highlight some of the essential issues of this complex field. Kraichnan's (1959) pioneering work on the DIA closure theory was followed by related, but independently developed, non-Markovian closures by Herring (1965) and McComb (1974, 1990). These so-called self-consistent field theory (SCFT) and local energy transfer theory (LET) closures have subsequently been shown to differ from the DIA only in how a fluctuation dissipation theorem is invoked (Frederiksen et al. 1994; Frederiksen and Davies 2000; Kiyani and McComb 2004). The SCFT and LET closures have very similar performance to the DIA for homogeneous turbulence at finite Reynolds number (Frederiksen et al. 1994; Frederiksen and Davies 2000).

Over the last two decades there has been increased interest in time-dependent non-equilibrium quantum field theories (Berges 2004; Calzetta and Hu 2008; Berges 2016) with applications to Bose-Einstein condensation far from equilibrium (Gasenzer 2009; Berges and Sexty 2012), cosmology and inflation (Kofman et al. 1997; Micha and Tkachev 2003; Kofman 2008) and quark-gluon plasma (Arnold 2007). In these problems it is necessary to allow for the evolution of quantum fluctuations in a background of dynamical spatially inhomogeneous fields. The statistical dynamical equations for these problems are typically formulated using the closed time path (CTP) formalism of Schwinger (1961) and Keldysh (1965). The relationship between the Schwinger-Keldysh CTP formalism for time-dependent quantum fields and the MSR formalism for classical fields was established by Cooper et al. (2001). They noted that the two formalisms gave the same statistical equations in the case of quadratic nonlinearity (cubic Lagrangian) but with extra vertices (and different initial conditions) in the quantum case. Berges and Ganzer (2007) also examined the relationships between the Schwinger-Keldysh CTP formalism for quantum systems and the MSR and Lagrangian path integral formalisms for classical systems; they compared the evolution of a classical Bose gas with the quantum counterpart. Some studies have used models in one space dimension (Aarts and Berges 2002; Cooper et al. 2003; Boyanovsky et al. 2004) or two space dimensions (Juchem et al. 2004) for the spatially homogeneous problem or in the case of the inhomogeneous problem for $\lambda\phi^4$ theory have used the simplified Hartree approximation (Bettencourt et al. 2001). Many studies have also focused on the classical limit of the quantum theories (Aarts and Smit 1998; Blagoev et al. 2001) which is restricted to the early stages of evolution (Berges, 2016). Blagoev et al. (2001) and Cooper et al. (2003) find that the DIA (or, their term, the BVA) performs better than the two particle irreducible (2PI-1/N) expansion and much better than the Hartree approximation. Some studies (Berges et al. 2003; Cooper et al. 2005; Arrizabalaga et al. 2005) have also considered three space dimensions for the spatially homogeneous problem with or without symmetry breaking and mean fields.

The development of general and elegant formalisms for describing the statistical dynamics of both quantum and classical fields has been an impressive achievement. However, the numerical solution of the above formulations of statistical dynamical equations for inhomogeneous fields in several dimensions poses severe computational challenges. This is the case even in the bare vertex approximation, such as for Kraichnan's (1964, 1972) inhomogeneous DIA (IDIA), which has not been numerically implemented; indeed at the time Kraichnan recognized that it was computationally intractable at any reasonable resolution.

A computationally tractable inhomogeneous non-Markovian closure theory, termed the quasi-diagonal direct interaction approximation (QDIA), was formulated by Frederiksen (1999) for two-dimensional turbulent flow over topography. This required a different method of deriving the renormalized statistical dynamical equations from the methods described above. In this alternative approach, the two-point functions are taken as homogeneous – diagonal in Fourier space – to zero order in perturbation theory and the first order off-diagonal elements of the covariance and response functions are expressed in terms of the diagonal elements and the mean-fields and topography. The first and second order statistical QDIA equations are then formally renormalized. The theory has also been generalized to inhomogeneous Rossby wave turbulence (Frederiksen and O'Kane 2005) and to general classical field theories with quadratic nonlinearity, including quasi-geostrophic (QG) baroclinic and three-dimensional inhomogeneous turbulence with general mean fields (Frederiksen 2012a, b). These equations are first order in the tendency or time-derivative.

The QDIA statistical closure is only a few times more computationally intensive than the homogeneous DIA unlike Kraichnan's (1964, 1972) inhomogeneous IDIA and related bare vertex approximation closures for inhomogeneous quantum fields. This has been established in statistical dynamical studies by O'Kane and Frederiksen (2004) and Frederiksen and O'Kane (2005) where the QDIA was implemented numerically for turbulent flows on an *f*-plane (non-rotating flows) and *β*-plane (differentially rotating flows) respectively. In particular, Frederiksen and O'Kane (2005) studied Rossby wave dispersion due to eastward zonal flows impinging on an isolated topographic feature, in a moderate Reynolds number turbulent environment; they found pattern correlations between the QDIA and an ensemble of 1800 direct numerical simulations to be as high as 0.9999 for the mean Rossby wave trains in 10 day simulations. This level of agreement is remarkable for this is a far from equilibrium process that severely tests the closure performance. The QDIA closure has been further extensively tested and applied to problems in predictability (Frederiksen and O'Kane 2005; O'Kane and Frederiksen 2008a), data assimilation (O'Kane and Frederiksen 2008b) and subgrid modeling (O'Kane and Frederiksen 2008c; Frederiksen and O'Kane 2008). The performance of the QDIA has been enhanced (O'Kane and Frederiksen 2004; Frederiksen and O'Kane 2005) through a cumulant update restart procedure (Rose 1985), that employs non-Gaussian initial conditions, as in earlier studies with homogeneous non-Markovian closures (Frederiksen et al. 1994; Frederiksen and Davies 2000, 2004). In the case of high Reynolds number turbulence, a regularized version of the QDIA employs a one parameter empirical vertex renormalization, as for homogeneous turbulence (Frederiksen and Davies 2004), that ensures the right power law behavior (O'Kane and Frederiksen 2004).

The aim of this article is to generalize the QDIA closure theory to equations that are second order in the tendency, or time-derivative, such as the Klein-Gordon equation and to incorporate non-Gaussian noise effects and quantum effects. The focus is on equations with quadratic nonlinearity in the field equations (cubic in the Lagrangian) although higher order interactions may also be covered such as for $\lambda\phi^4$ Lagrangian theories in the auxiliary field formulation (Blagoev et al. 2001; Bender et al. 1977).

The plan of this paper is as follows. In Section 2 we document the general form of the stochastic differential equations, which are first or second order in the tendency, that are considered in the article. In the case of classical fluid dynamics, specific examples are for geophysical turbulent flows described by the quasi-geostrophic equations and for three dimensional turbulence described by the Navier Stokes equations. As well, the second order tendency Klein Gordon equations for classical and quantum fields, interacting through a $g\phi^3$ Lagrangian term, and coupled field-auxiliary field equations associated with a $\lambda\phi^4$ Lagrangian, are considered. The general form of the dynamical equations in Fourier (momentum) space is also presented. The inhomogeneous IDIA closure equations that correspond to a bare vertex approximation are formulated in Section 3 and the inclusion of non-Gaussian noise effects is considered in Section 4. In Section 5, the inhomogeneous Self-Energy (SE) closure equations are derived as modifications of the IDIA that are second order in the interaction coefficient (coupling constant) in both the mean field equation as well as in the two-point equations for the covariance and response function. In Section 6 the more computationally efficient QDIA closure equations are obtained from the Self-Energy closure by assuming that to lowest order the fields are homogeneous or diagonal in Fourier space; the off-diagonal components of the two-point functions are expressed in terms of the diagonal components and the mean-field and possibly topography. The full QDIA closure equations for the particular example of scalar fields interacting through a $g\phi^3$ Lagrangian term are presented in Section 7. There the substitutions required for obtaining the QDIA closure, for the coupled field-auxiliary field equations associated with a $\lambda\phi^4$ Lagrangian, from the general theory in Section 6 are also presented. In Section 8 we summarize the conclusions and discuss applications and restart procedures. The functional equations of MSR and the path integral approach of Jensen (1981) are summarized in Appendix A. There the inclusion of non-Gaussian noise and non-Gaussian initial conditions is also considered. In Appendix B a derivation relating the first order inhomogeneous elements of the two-point functions to the mean-field and topography, needed for the Self-Energy and QDIA closures, is presented.

## 2. Equations for Inhomogeneous Classical and Quantum Fields

The statistical closure equations formulated in this paper apply to a wide variety of classical and quantum field theories with quadratic nonlinearity. We consider stochastic differential equations that are first or second order in the tendency of the form:

$$\frac{\partial^n}{\partial t_1^n}\psi(1) = U_1(1) + U_2(1,2)\psi(2) + U_3(1,2,3)\psi(2)\psi(3) \qquad (2.1)$$

with $n = 1$ or $2$. Here $\psi(1)$ is a multicomponent field with initial condition $\psi_0(1)$ and $1 = (\mathbf{x}_1, a_1, t_1) = (\mathbf{1}, t_1)$ is a short hand notation for the time $(t_1)$, space or spectral space $(\mathbf{x}_1)$, and indices $(a_1)$ representing for example different fields or fields at different levels. First order tendency equations of the form (2.1) were considered by MSR and Jensen (1981). In the path integral formalism of Jensen (1981) the variables $U_i(1,...,i) = \overline{U}_i(1,...,i) + \tilde{U}_i(1,...,i)$ consist of a mean deterministic part $\overline{U}_i(1,...,i)$ and a random part $\tilde{U}_i(1,...,i)$. We assume without loss of generality that $U_3(1,2,3) = U_3(1,3,2)$. As well $\psi_0(1) = \overline{\psi}_0(1) + \tilde{\psi}_0(1)$ where again $\overline{\psi}_0(1)$ is deterministic and $\tilde{\psi}_0(1)$ is random. Summation over repeated discrete indices and integration over repeated continuous variables is assumed.

Next we present a few examples of interest in fluid dynamics and non-equilibrium quantum field theories.

*2.1. Classical fields described by first order tendency equations*

Kraichnan's (1959, 1964, 1972) homogeneous DIA and inhomogeneous IDIA, and related statistical closures, and the MSR formalism and classical path integral formalisms (Phythian 1977; Jensen 1981) have focused on first order tendency equations for classical fluid dynamics. Of particular interest have been the equations for Navier Stokes turbulence in two and three dimensions and the quasi-geostrophic equations for geophysical fluid dynamics.

*2.1a. Quasi-geostrophic equations for turbulent flow*

Taking suitable length and time scales, the non-dimensional equation for 2-level baroclinic quasi-geostrophic flow over topography on an *f*-plane may be written in the form:

$$\frac{\partial q^a}{\partial t} = -J(\psi^a, q^a + h^a) - D_0^{ab} q^b + f_0^a. \qquad (2.2a)$$

Here, $a = 1$ or $2$, $\psi^a$ is the streamfunction and $q^a = \nabla^2 \psi^a + (-1)^a F_L(\psi^1 - \psi^2)$ is the reduced potential vorticity, $\omega^a = \nabla^2 \psi^a$ is the relative vorticity, $h^2 = h$, $h^1 = 0$ where $h$ is the scaled topography, $D_0^{ab}$ are dissipation operators and $f_0^a$ are forcing functions. Also, $F_L$ is the layer coupling parameter (Frederiksen 2012a) which is inversely proportional to the static stability. In planar geometry:

$$J(\psi,\zeta) = \frac{\partial \psi}{\partial x}\frac{\partial \zeta}{\partial y} - \frac{\partial \psi}{\partial y}\frac{\partial \zeta}{\partial x} \tag{2.2b}$$

where $\mathbf{x} = (x, y)$ is the position and $t$ is time.

*2.1b. Navier Stokes equations for three-dimensional turbulent flow*

The Navies Stokes equations for three-dimensional inhomogeneous turbulent flow may be written in the form (McComb 1990):

$$\frac{\partial}{\partial t}u^a + u^\beta \frac{\partial u^a}{\partial x^\beta} = -\frac{1}{\rho}\frac{\partial p}{\partial x^a} + \nu_0 \nabla^2 u^a + f_0^a. \tag{2.3}$$

Here, $u^a(\mathbf{x},t)$, $a=1,2,3$ is the fluid velocity at position $\mathbf{x}$ and time $t$. Also, $\rho$ is the density and $p(\mathbf{x},t)$ is the pressure. The prescribed viscosity is specified by $\nu_0$, and $f_0^a(\mathbf{x},t)$ are forcing functions.

*2.1c. General form of first order tendency equations*

The quasi-geostrophic equations for geophysical flow and the Navier Stokes equations are examples of first order tendency classical equations that are quadratic in the field variables and for which the MSR and classical path integral formalisms may be applied directly to yield statistical dynamical equations. MSR write their equations for the field and its adjoint as in Eq. (A.5) or in component form:

$$\begin{aligned}\frac{\partial}{\partial t_1}\psi(1) &= \gamma_1^2(1) + \gamma_2^{21}(1,2)\psi(2) + \gamma_2^{22}(1,2)\hat{\psi}(2) + \tfrac{1}{2}\gamma_3^{211}(1,2,3)\psi(2)\psi(3) \\ &+ \tfrac{1}{2}\gamma_3^{212}(1,2,3)\psi(2)\hat{\psi}(3) + \tfrac{1}{2}\gamma_3^{221}(1,2,3)\hat{\psi}(2)\psi(3) + \tfrac{1}{2}\gamma_3^{222}(1,2,3)\hat{\psi}(2)\hat{\psi}(3)\end{aligned} \tag{2.4}$$

with adjoint equation

$$\begin{aligned}-\frac{\partial}{\partial t_1}\hat{\psi}(1) &= \gamma_1^1(1) + \gamma_2^{12}(1,2)\hat{\psi}(2) + \gamma_2^{11}(1,2)\psi(2) + \tfrac{1}{2}\gamma_3^{122}(1,2,3)\hat{\psi}(2)\hat{\psi}(3) \\ &+ \tfrac{1}{2}\gamma_3^{121}(1,2,3)\hat{\psi}(2)\psi(3) + \tfrac{1}{2}\gamma_3^{112}(1,2,3)\psi(2)\hat{\psi}(3) + \tfrac{1}{2}\gamma_3^{111}(1,2,3)\psi(2)\psi(3).\end{aligned} \tag{2.5}$$

Here $\psi(1)$ is the dynamical field and $\hat{\psi}(1)$ the adjoint. These equations are further described in Appendix A where the relationships between the $U$ and $\gamma$ coefficients are also discussed. In particular, the vertex $\gamma_3^{222}$ accounts for non-Gaussian noise or quantum effects. As also noted in Appendix A, for $\gamma_1$, $\gamma_2$ and $\gamma_3$ the first superscript refers to the conjugate spinor field while the later superscripts refer to the correct spinor field. This choice within the MSR and Jensen (1981) formalisms, related to the Pauli matrices (Eq. (A.2)), should be noted as it is somewhat counter intuitive. Examples of the parameters $\gamma_1$, $\gamma_2$ and $\gamma_3$ for three-dimensional Navier Stokes flows and quasi-geostrophic turbulence are given in Frederiksen (2012a, b).

*2.2. Second-order tendency Klein-Gordon equation for classical and quantum fields*

Cooper et al. (2001) show that the first order tendency MSR and classical path integral formalisms can also be applied to second-order tendency, or time-derivative, equations like the Klein-Gordon equation. This is done by rewriting the second-order tendency equations in terms of the fields $\phi(1)$ and the canonical momentum fields $\pi(1) = \partial \phi(1)/\partial t_1$. They also show that the MSR and path integral formalisms can be written directly in a covariant second-order tendency form and are related to the Schwinger-Keldysh closed time path (CTP) formalism for quantum dynamics (apart from initial conditions and extra vertices for quantum effects).

*2.2a. Lagrangian with interaction $g\phi^3$*

The MSR and path integral formalisms can be applied directly in second-order in time covariant form for general $g\phi^3$ theories as noted by Cooper et al. (2001). The only difference in the associated statistical equations are the addition of extra vertices for the quantum case based on the CTP formalism. In this respect the statistical equations for quantum field theories are analogous to those of classical field theories with non-Gaussian noise, as we explore in more detail. Cooper et al. (2001, Eq. (38)) give the Lagrangian for a multi-field model of the $g\phi^3$ theory in second-order tendency form. Their formalism has the time-derivative and nonlinear term on the same side of the equation, unlike our Eq. (2.1). For our parallel development of the statistical dynamical equations, for first and second-order tendency equations, it is however more convenient to keep the form (2.1), which changes the sign of the $U_3$ and $\gamma_3$ vertices from those of Cooper et al. (2001) and Blagoev et al. (2001). In terms of the field $\phi(1)$ and the adjoint of the canonical momentum field $\hat{\pi}(1)$ our dynamical equations are:

$$\frac{\partial^2}{\partial t_1^2} \phi(1) = \gamma_1^2(1) + \gamma_2^{21}(1,2)\phi(2) + \gamma_2^{22}(1,2)\hat{\pi}(2) + \tfrac{1}{2}\gamma_3^{211}(1,2,3)\phi(2)\phi(3)$$
$$+ \tfrac{1}{2}\gamma_3^{212}(1,2,3)\phi(2)\hat{\pi}(3) + \tfrac{1}{2}\gamma_3^{221}(1,2,3)\hat{\pi}(2)\phi(3) + \tfrac{1}{2}\gamma_3^{222}(1,2,3)\hat{\pi}(2)\hat{\pi}(3)$$
(2.6)

and

$$\frac{\partial^2}{\partial t_1^2} \hat{\pi}(1) = \gamma_1^1(1) + \gamma_2^{12}(1,2)\hat{\pi}(2) + \gamma_2^{11}(1,2)\phi(2) + \tfrac{1}{2}\gamma_3^{122}(1,2,3)\hat{\pi}(2)\hat{\pi}(3)$$
$$+ \tfrac{1}{2}\gamma_3^{121}(1,2,3)\hat{\pi}(2)\phi(3) + \tfrac{1}{2}\gamma_3^{112}(1,2,3)\phi(2)\hat{\pi}(3) + \tfrac{1}{2}\gamma_3^{111}(1,2,3)\phi(2)\phi(3).$$
(2.7)

The formulation of these equations is further documented in Appendix A. For example, in the case of a single quantum field, the dynamical equations, obtained from the Schwinger-Keldysh CTP formalism (Schwinger 1961; Keldysh 1965) can be written in the form:

$$\left[\frac{\partial^2}{\partial t^2} - \nabla^2 + m^2\right]\phi + \tfrac{1}{2}g\phi^2 + \tfrac{1}{8}g\hbar^2\hat{\pi}_\phi^2 = j^\phi,$$
$$\left[\frac{\partial^2}{\partial t^2} - \nabla^2 + m^2\right]\hat{\pi}_\phi + g\phi\hat{\pi}_\phi = j^{\hat{\pi}_\phi}.$$
(2.8)

Here, the right hand sides represent sources, $m$ is the mass and $g$ is the coupling constant with

$$g\delta(1-2)\delta(1-3) = -\gamma_3^{211}(1,2,3) = -\gamma_3^{121}(1,2,3) = -\gamma_3^{122}(1,2,3).$$
(2.9a)

Also, we note that the classical equations are the same as the above quantum equations apart from the term $\tfrac{1}{8}g\hbar^2$, where $\hbar$ is Planck's constant (over $2\pi$) with

$$\tfrac{1}{4} g\hbar^2 \delta(1-2)\delta(1-3) = -\gamma_3^{222}(1,2,3). \tag{2.9b}$$

The other three-point vertices in Eqs. (2.6) and (2.7) vanish. Here

$$\delta(1-2) = \delta_{a_1 a_2}\delta(\mathbf{x}_1 - \mathbf{x}_2)\delta(t_1 - t_2) \tag{2.9c}$$

where $\delta$ denotes the Kronecker delta function for discrete indices and the Dirac delta function for continuous variables. Of course the initial conditions for classical and quantum systems will differ (Cooper et al. 2001). The QDIA statistical closure equations for this $g\phi^3$ field theory are detailed in Section 7.1.

*2.2b. Coupled field-auxiliary field equations associated with $\lambda\phi^4$ Lagrangian interaction*

Blagoev et al. (2001) consider dynamical equations for coupled quadratic interactions (cubic Lagrangian) between a scalar field $\phi$ and an auxiliary field $\chi$ that can be related to the cubic self-interaction of $\phi$ within a $\lambda\phi^4$ Lagrangian interaction theory. In their auxiliary field formulation the equations for a single field $\phi$ and auxiliary field $\chi$ take the form:

$$\begin{aligned}
\left[\frac{\partial^2}{\partial t^2} - \nabla^2 + m^2\right]\phi + g\chi\phi + \tfrac{1}{4}g\hbar^2 \hat{\pi}_\phi \hat{\pi}_\chi &= j^\phi, \\
\left[\frac{\partial^2}{\partial t^2} - \nabla^2 + M^2\right]\chi + \tfrac{1}{2}g\phi^2 + \tfrac{1}{8}g\hbar^2 \hat{\pi}_\phi^2 &= j^\chi, \\
\left[\frac{\partial^2}{\partial t^2} - \nabla^2 + m^2\right]\hat{\pi}_\phi + g\chi\hat{\pi}_\phi + g\phi\hat{\pi}_\chi &= j^{\hat{\pi}_\phi}, \\
\left[\frac{\partial^2}{\partial t^2} - \nabla^2 + M^2\right]\hat{\pi}_\chi + g\phi\hat{\pi}_\phi &= j^{\hat{\pi}_\chi}.
\end{aligned} \tag{2.10a}$$

Again, the right hand sides represent sources, $g$ is a coupling parameter, $m$ and $M$ are mass parameters. The classical equations are the same as the above quantum equations apart from the terms, proportional to $g\hbar^2$. As also noted by Blagoev et al. (2001), after determining statistical closure equations associated with Eq. (2.10a) the composite limit (their Eq. (A.24) with $j \equiv j^\phi, S \equiv j^\chi$ but with $m^2 \to 0$ missing) must be taken. That is

$$\chi(\mathbf{x},t) \to \mu^2 + \tfrac{1}{2}\lambda\phi^2(\mathbf{x},t), \tag{2.10b}$$
$$m^2 \to 0; M^2 \to -\lambda^{-1}; g \to 1; j^\chi \to \mu^2\lambda^{-1}. \tag{2.10c}$$

As well, the bare retarded propagator or response function for the $\chi$ field is replaced by $-\lambda\delta(\mathbf{x}-\mathbf{x}')\delta(t-t')$. Here $\mu$ is the mass and $\lambda$ is the coupling constant for the $\lambda\phi^4$ theory. These equations can also be generalized to multiple $\phi$ fields (Cooper et al. 2003). With $\phi^a(\mathbf{x},t) = [\phi^1(\mathbf{x},t),...,\phi^N(\mathbf{x},t),\chi(\mathbf{x},t)]$ the field equations again take the form in Equations (2.6) and (2.7).

The scope of the current study is to focus on equations that are quadratic in the interactions with cubic interactions (quartic Lagrangian) for consideration in future work. However, we shall use the auxiliary field equations as an example of the more complex structure of the QDIA inhomogeneous closure equations when several fields are coupled; this is discussed in Section 7.2.

*2.3. Dynamical equations in spectral space*

The dynamical equations summarized in this section, and the statistical equations of Appendix A, can equally be formulated in Fourier or other spectral spaces (Frederiksen 2012b and references therein). Here we consider the discrete Fourier transform of the equations but the continuum Fourier transform (Kraichnan 1959; Carnevale and Frederiksen 1983) could equally be employed. In spectral space the equation corresponding to Eq. (2.1) can be obtained through the Fourier transforms:

$$\psi(1) \equiv \psi^{a_1}(\mathbf{x}_1, t_1) = \sum_{\mathbf{k}_1} \zeta^{a_1}_{\mathbf{k}_1}(t) \exp(i\mathbf{k}_1 \cdot \mathbf{x}_1) \tag{2.11a}$$

where

$$\zeta^{a}_{\mathbf{k}}(t) = \frac{1}{(2\pi)^d} \int_o^{2\pi} d^d \mathbf{x} \, \psi^a(\mathbf{x}, t) \exp(-i\mathbf{k} \cdot \mathbf{x}), \tag{2.11b}$$

and $d$ is the dimension; for example, the position $\mathbf{x} = (x, y, z)$ and wave number $\mathbf{k} = (k_x, k_y, k_z)$ for the case when the dimension $d = 3$. Here, we have assumed that suitable length scales are used so that the domain in Eq. (2.11b) is between $0$ and $2\pi$. The wave number cut-offs in Eq. (2.11a) are to be specified in simulations and statistical calculations and need to be suitably chosen together with the renormalized parameters like viscosities, forcing, masses etc. as discussed in Section 8.

We begin by considering a set of general classical dynamical equations in spectral form:

$$\frac{\partial^n}{\partial t^n} \zeta^a_{\mathbf{k}}(t) + \sum_{\mathbf{k}'} D_0^{\alpha\beta}(\mathbf{k}, \mathbf{k}') \zeta^{\beta}_{\mathbf{k}'}(t) \\ = \sum_{\mathbf{p}} \sum_{\mathbf{q}} \delta(\mathbf{k}, \mathbf{p}, \mathbf{q}) \left[ K^{abc}(\mathbf{k}, \mathbf{p}, \mathbf{q}) \zeta^b_{-\mathbf{p}}(t) \zeta^c_{-\mathbf{q}}(t) + A^{abc}(\mathbf{k}, \mathbf{p}, \mathbf{q}) \zeta^b_{-\mathbf{p}}(t) h^c_{-\mathbf{q}} \right] + f_0^a(\mathbf{k}, t). \tag{2.12}$$

The associated adjoint equations and the roles of 'spurious' vertices (Rose 1974; Jensen 1981) for non-Gaussian and quantum effects are documented in Appendix A and in Sections 4 and 7 below. In Eq. (2.12), we have made the replacements $U_1(1) \to f_0^a(\mathbf{k}, t)$, $U_3(1,2,3) \to K^{abc}(\mathbf{k}, \mathbf{p}, \mathbf{q})$ and $U_2(1,2)$ has resulted in the terms $-D_0^{\alpha\beta}(\mathbf{k}, \mathbf{k}')$ and $A^{abc}(\mathbf{k}, \mathbf{p}, \mathbf{q}) h^c_{-\mathbf{q}}$ to include the possibility of topography $h^c_{-\mathbf{q}}$ in the classical fluid case. We also note that

$$\zeta^a_{-\mathbf{k}} = \zeta^{a*}_{\mathbf{k}}, \tag{2.13}$$

$$K^{abc}(\mathbf{k}, \mathbf{p}, \mathbf{q}) = K^{acb}(\mathbf{k}, \mathbf{q}, \mathbf{p}) \tag{2.14}$$

$$\delta(\mathbf{k}, \mathbf{p}, \mathbf{q}) = \begin{cases} 1 \text{ if } \mathbf{k} + \mathbf{p} + \mathbf{q} = 0 \\ 0 \text{ otherwise.} \end{cases} \tag{2.15}$$

Our interest is in the case where

$$D_0^{\alpha\beta}(\mathbf{k}, \mathbf{k}') = D_0^{\alpha\beta}(\mathbf{k}) \delta_{\mathbf{k}, \mathbf{k}'} \tag{2.16}$$

with $\delta_{\mathbf{k}, \mathbf{k}'}$ the Kronecker delta function. Thus $D_0^{\alpha\beta}(\mathbf{k}, \mathbf{k}') = D_0^{\alpha\beta}(\mathbf{k})$ when $\mathbf{k}' = \mathbf{k}$ and is zero otherwise. We have introduced the more general form in Equation (2.16) for later convenience. The interaction coefficients $A^{abc}(\mathbf{k},\mathbf{p},\mathbf{q})$ and $K^{abc}(\mathbf{k},\mathbf{p},\mathbf{q})$ for Navier Stokes and quasi-geostrophic flows are documented in Frederiksen (2012a, b); for these flows the dissipation $D_0^{\alpha\beta}(\mathbf{k})$ and the viscosity $\nu_0^{\alpha\beta}(\mathbf{k})$ are related through $D_0^{\alpha\beta}(\mathbf{k}) = \nu_0^{\alpha\beta}(\mathbf{k})\mathbf{k}^2$.

## 3. Inhomogeneous IDIA Closure Equations

Kraichnan's (1964, 1972) inhomogeneous IDIA statistical equations for classical fields were derived through a formal series reversion approach before being put on a firmer foundation by the MSR formalism and the path integral formalism (Phythian 1977; Jensen 1981). They are essentially the equations for the mean flow and a measure of the fluctuations (two-point function) obtained by Reynolds averaging (McComb 1990). The IDIA closes the equations by expressing the three-point function in terms of the two-point functions. A summary of the derivation following Jensen (1981) is given in Appendix A including the effects of non-Gaussian noise and non-Gaussian initial conditions. Initially, we postpone consideration of non-Gaussian noise effects and related quantum effects until Section 4. We also represent the statistical equations in Fourier space for comparison with the much more efficient QDIA closure equations in Section 6. The QDIA equations were derived directly (Frederiksen 1999, 2012a) for first order tendency equations but it is probably more enlightening to see how, in Section 5, the IDIA equations can be modified to form the Self-Energy closure equations (Frederiksen 2012b) and from these the QDIA emerge through diagonal dominance in spectral space in Section 6. The IDIA, Self-Energy and QDIA statistical closure equations are realizable with underpinning generalized Langevin equations (Frederiksen 2012a, b and references therein); the same applies for second-order tendency equations.

We now consider an ensemble of simulations and express the field component for a given realization by:

$$\zeta_\mathbf{k}^a = <\zeta_\mathbf{k}^a> + \tilde{\zeta}_\mathbf{k}^a \tag{3.1}$$

where $<\zeta_\mathbf{k}^a>$ is the ensemble mean and $\tilde{\zeta}_\mathbf{k}^a$ denotes the deviation from the ensemble mean. The spectral equation (2.12) can then be expressed in terms of $<\zeta_\mathbf{k}^a>$ and $\tilde{\zeta}_\mathbf{k}^a$ as follows:

$$\frac{\partial^n}{\partial t^n}<\zeta_\mathbf{k}^a(t)> + \sum_{\mathbf{k}'} D_0^{\alpha\beta}(\mathbf{k},\mathbf{k}')<\zeta_{\mathbf{k}'}^\beta(t)> = \\
\sum_\mathbf{p}\sum_\mathbf{q} \delta(\mathbf{k},\mathbf{p},\mathbf{q})[K^{abc}(\mathbf{k},\mathbf{p},\mathbf{q})\{<\zeta_{-\mathbf{p}}^b(t)><\zeta_{-\mathbf{q}}^c(t)> + C_{-\mathbf{p},-\mathbf{q}}^{bc}(t,t)\} \\
+ A^{abc}(\mathbf{k},\mathbf{p},\mathbf{q})<\zeta_{-\mathbf{p}}^b(t)> h_{-\mathbf{q}}^c] + \bar{f}_0^a(\mathbf{k},t), \tag{3.2}$$

$$\frac{\partial^n}{\partial t^n}\tilde{\zeta}_\mathbf{k}^a(t) + \sum_{\mathbf{k}'} D_0^{\alpha\beta}(\mathbf{k},\mathbf{k}')\tilde{\zeta}_{\mathbf{k}'}^\beta(t) = \\
\sum_\mathbf{p}\sum_\mathbf{q} \delta(\mathbf{k},\mathbf{p},\mathbf{q})[K^{abc}(\mathbf{k},\mathbf{p},\mathbf{q})\{<\zeta_{-\mathbf{p}}^b(t)>\tilde{\zeta}_{-\mathbf{q}}^c(t) + \tilde{\zeta}_{-\mathbf{p}}^b(t)<\zeta_{-\mathbf{q}}^c(t)> \\
+ \tilde{\zeta}_{-\mathbf{p}}^b(t)\tilde{\zeta}_{-\mathbf{q}}^c(t) - C_{-\mathbf{p},-\mathbf{q}}^{bc}(t,t)\} + A^{abc}(\mathbf{k},\mathbf{p},\mathbf{q})\tilde{\zeta}_{-\mathbf{p}}^b(t)h_{-\mathbf{q}}^c] + \tilde{f}_0^a(\mathbf{k},t). \tag{3.3}$$

Here:

$$f_0^a(\mathbf{k}) = \bar{f}_0^a(\mathbf{k}) + \tilde{f}_0^a(\mathbf{k}), \tag{3.4}$$

$$\bar{f}_0^a(\mathbf{k}) = <f_0^a(\mathbf{k})>, \tag{3.5}$$

$$C_{-\mathbf{p},-\mathbf{q}}^{bc}(t,s) = <\tilde{\zeta}_{-\mathbf{p}}^b(t)\tilde{\zeta}_{-\mathbf{q}}^c(s)>, \tag{3.6}$$

are the mean and random forcing functions and two-time covariance matrix elements.

### 3.1. Statistical closure equations

For closure, the mean-field equation, Eq. (3.2) or Eq. (A.24), requires knowledge of the single-time covariance $C_{-\mathbf{p},-\mathbf{q}}^{bc}(t,t)$. In turn, the equation for the covariance requires knowledge of the three-point cumulant (Eq. (A.12)). This may be seen by simply multiplying the terms in Eq. (3.3) by $\tilde{\zeta}_{-\mathbf{l}}^{\alpha}(t')$ and taking the statistical average:

$$\frac{\partial^n}{\partial t^n} C_{\mathbf{k},-\mathbf{l}}^{a\alpha}(t,t') + \sum_{\mathbf{k}'} D_0^{a\beta}(\mathbf{k},\mathbf{k}')C_{\mathbf{k}',-\mathbf{l}}^{\beta\alpha}(t,t') = \sum_{\mathbf{p}}\sum_{\mathbf{q}} \delta(\mathbf{k},\mathbf{p},\mathbf{q})A^{abc}(\mathbf{k},\mathbf{p},\mathbf{q})C_{-\mathbf{p},-\mathbf{l}}^{b\alpha}(t,t')h_{-\mathbf{q}}^c \tag{3.7}$$

$$+ \sum_{\mathbf{p}}\sum_{\mathbf{q}} \delta(\mathbf{k},\mathbf{p},\mathbf{q})K^{abc}(\mathbf{k},\mathbf{p},\mathbf{q})[<\zeta_{-\mathbf{p}}^b(t)>C_{-\mathbf{q},-\mathbf{l}}^{c\alpha}(t,t') + C_{-\mathbf{p},-\mathbf{l}}^{b\alpha}(t,t')<\zeta_{-\mathbf{q}}^c(t)>$$

$$+ <\tilde{\zeta}_{-\mathbf{p}}^b(t)\tilde{\zeta}_{-\mathbf{q}}^c(t)\tilde{\zeta}_{-\mathbf{l}}^{\alpha}(t')>] + \int_{t_o}^{t'} ds \sum_{\mathbf{k}'} F_0^{a\beta}(\mathbf{k},-\mathbf{k}',t,s)R_{-\mathbf{l},-\mathbf{k}'}^{\alpha\beta}(t',s).$$

This equation applies for $t > t'$ while for $t < t'$

$$C_{\mathbf{k},-\mathbf{l}}^{a\alpha}(t,t') = C_{-\mathbf{l},\mathbf{k}}^{\alpha a}(t',t). \tag{3.8}$$

The three-point function may be expressed in terms of the self-energies and two-point functions (Eq. (A.13a)) and in the DIA (Eq. (A.15)) takes the form:

$$\sum_{\mathbf{p}}\sum_{\mathbf{q}} \delta(\mathbf{k},\mathbf{p},\mathbf{q})K^{abc}(\mathbf{k},\mathbf{p},\mathbf{q}) < \tilde{\zeta}_{-\mathbf{p}}^b(t)\tilde{\zeta}_{-\mathbf{q}}^c(t)\tilde{\zeta}_{-\mathbf{l}}^{\alpha}(t') > \tag{3.9}$$

$$= \int_{t_o}^{t'} ds \sum_{\mathbf{k}'} S_{\mathbf{k},-\mathbf{k}'}^{a\beta}(t,s)R_{-\mathbf{l},-\mathbf{k}'}^{\alpha\beta}(t',s) - \int_{t_o}^{t} ds \sum_{\mathbf{k}'} \eta_{\mathbf{k},\mathbf{k}'}^{a\beta}(t,s)C_{-\mathbf{l},\mathbf{k}'}^{\alpha\beta}(t',s)$$

for Gaussian initial conditions. Non-Gaussian initial conditions are also considered in Appendix A, Section 9.3; for the sake of brevity we do not detail the non-Gaussian initial conditions here in spectral form but refer the interested reader to the articles by Rose (1985), Frederiksen et al. (1994), O'Kane and Frederiksen (2004) and Frederiksen and O'Kane (2005).

Thus, as in Eqs. (A.17) and (A.27), for $t > t'$ we obtain the closure

$$\frac{\partial^n}{\partial t^n} C_{\mathbf{k},-\mathbf{l}}^{a\alpha}(t,t') + \sum_{\mathbf{k}'} D_0^{a\beta}(\mathbf{k},\mathbf{k}')C_{\mathbf{k}',-\mathbf{l}}^{\beta\alpha}(t,t') = \sum_{\mathbf{p}}\sum_{\mathbf{q}} \delta(\mathbf{k},\mathbf{p},\mathbf{q})A^{abc}(\mathbf{k},\mathbf{p},\mathbf{q})C_{-\mathbf{p},-\mathbf{l}}^{b\alpha}(t,t')h_{-\mathbf{q}}^c \tag{3.10}$$

$$+ \sum_{\mathbf{p}}\sum_{\mathbf{q}} \delta(\mathbf{k},\mathbf{p},\mathbf{q})K^{abc}(\mathbf{k},\mathbf{p},\mathbf{q})[<\zeta_{-\mathbf{p}}^b(t)>C_{-\mathbf{q},-\mathbf{l}}^{c\alpha}(t,t') + C_{-\mathbf{p},-\mathbf{l}}^{b\alpha}(t,t')<\zeta_{-\mathbf{q}}^c(t)>]$$

$$+ \int_{t_o}^{t'} ds \sum_{\mathbf{k}'} N_{\mathbf{k},-\mathbf{k}'}^{a\beta}(t,s)R_{-\mathbf{l},-\mathbf{k}'}^{\alpha\beta}(t',s) - \int_{t_o}^{t} ds \sum_{\mathbf{k}'} \eta_{\mathbf{k},\mathbf{k}'}^{a\beta}(t,s)C_{-\mathbf{l},\mathbf{k}'}^{\alpha\beta}(t',s).$$

From Eq. (A.22), the nonlinear damping self-energy

$$\eta^{a\alpha}_{\mathbf{k},\mathbf{k}'}(t,s) = \tag{3.11}$$
$$-4\sum_{\mathbf{p}}\sum_{\mathbf{q}}\sum_{\mathbf{p}'}\sum_{\mathbf{q}'} \delta(\mathbf{k},\mathbf{p},\mathbf{q})\delta(\mathbf{k}',\mathbf{p}',\mathbf{q}')K^{abc}(\mathbf{k},\mathbf{p},\mathbf{q})K^{\beta\gamma\alpha}(-\mathbf{p}',-\mathbf{q}',-\mathbf{k}')R^{b\beta}_{-\mathbf{p},-\mathbf{p}'}(t,s)C^{c\gamma}_{-\mathbf{q},\mathbf{q}'}(t,s)$$
$$\equiv -{}^{\eta}\Sigma^{a\alpha}_{\mathbf{k},\mathbf{k}'}(t,s).$$

Also, from Eq. (A.23), the nonlinear noise

$$N^{a\alpha}_{\mathbf{k},-\mathbf{k}'}(t,s) = S^{a\alpha}_{\mathbf{k},-\mathbf{k}'}(t,s) + F^{a\alpha}_0(\mathbf{k},-\mathbf{k}',t,s) \tag{3.12}$$

where the nonlinear noise self-energy

$$S^{a\alpha}_{\mathbf{k},-\mathbf{k}'}(t,s) = \tag{3.13}$$
$$2\sum_{\mathbf{p}}\sum_{\mathbf{q}}\sum_{\mathbf{p}'}\sum_{\mathbf{q}'} \delta(\mathbf{k},\mathbf{p},\mathbf{q})\delta(\mathbf{k}',\mathbf{p}',\mathbf{q}')K^{abc}(\mathbf{k},\mathbf{p},\mathbf{q})K^{\alpha\beta\gamma}(-\mathbf{k}',-\mathbf{p}',-\mathbf{q}')C^{b\beta}_{-\mathbf{p},\mathbf{p}'}(t,s)C^{c\gamma}_{-\mathbf{q},\mathbf{q}'}(t,s)$$
$$\equiv {}^{S}\Sigma^{a\alpha}_{\mathbf{k},-\mathbf{k}'}(t,s)$$

and the covariance of the prescribed noise or random forcing

$$F^{a\alpha}_0(\mathbf{k},-\mathbf{k}',t,s) = <\tilde{f}^a_0(\mathbf{k},t)\tilde{f}^\alpha_0(-\mathbf{k}',s)>. \tag{3.14a}$$

We assume homogeneous random forcing so that

$$F^{a\beta}_0(\mathbf{k},-\mathbf{k}',t,s) = F^{a\beta}_0(\mathbf{k},-\mathbf{k},t,s)\delta_{\mathbf{k},\mathbf{k}'} \tag{3.14b}$$

where $\delta_{\mathbf{k},\mathbf{k}'}$ is the Kronecker delta function. Thus $F^{a\beta}_0(\mathbf{k},-\mathbf{k}',t,s) = F^{a\beta}_0(\mathbf{k},-\mathbf{k},t,s)$ when $\mathbf{k}'=\mathbf{k}$ and is zero otherwise. We have introduced the more general form for later convenience.

The equation for the (retarded) response function is given in Eq. (A.18). That is, for $t > t'$

$$\frac{\partial^n}{\partial t^n}R^{a\alpha}_{\mathbf{k},\mathbf{l}}(t,t') + \sum_{\mathbf{k}'}D^{a\beta}_0(\mathbf{k},\mathbf{k}')R^{\beta\alpha}_{\mathbf{k}',\mathbf{l}}(t,t') = \sum_{\mathbf{p}}\sum_{\mathbf{q}}\delta(\mathbf{k},\mathbf{p},\mathbf{q})A^{abc}(\mathbf{k},\mathbf{p},\mathbf{q})R^{b\alpha}_{-\mathbf{p},\mathbf{l}}(t,t')h^c_{-\mathbf{q}} \tag{3.15}$$
$$+ \sum_{\mathbf{p}}\sum_{\mathbf{q}}\delta(\mathbf{k},\mathbf{p},\mathbf{q})K^{abc}(\mathbf{k},\mathbf{p},\mathbf{q})[<\zeta^b_{-\mathbf{p}}(t)>R^{c\alpha}_{-\mathbf{q},\mathbf{l}}(t,t') + R^{b\alpha}_{-\mathbf{p},\mathbf{l}}(t,t')<\zeta^c_{-\mathbf{q}}(t)>]$$
$$-\int_{t'}^t ds \sum_{\mathbf{k}'} \eta^{a\beta}_{\mathbf{k},\mathbf{k}'}(t,s)R^{\beta\alpha}_{\mathbf{k}',\mathbf{l}}(s,t').$$

Here, $R^{ab}_{\mathbf{k},\mathbf{l}}(t,t) = \delta_{\mathbf{k},\mathbf{l}}\delta^{ab}$ and $\delta$ is the Kronecker delta function.

The singe-time cumulant equation may be obtained from the expression

$$\frac{\partial^n}{\partial t^n}C^{ab}_{\mathbf{k},-\mathbf{l}}(t,t) = \lim_{t' \to t}\{\frac{\partial^n}{\partial t^n}C^{ab}_{\mathbf{k},-\mathbf{l}}(t,t') + \frac{\partial^n}{\partial t'^n}C^{ab}_{\mathbf{k},-\mathbf{l}}(t,t')\} \tag{3.16}$$
$$= \lim_{t' \to t}\{\frac{\partial^n}{\partial t^n}C^{ab}_{\mathbf{k},-\mathbf{l}}(t,t') + \frac{\partial^n}{\partial t'^n}C^{ba}_{-\mathbf{l},\mathbf{k}}(t',t)\}.$$

This yields

$$\frac{\partial^n}{\partial t^n} C^{a\alpha}_{\mathbf{k},-\mathbf{l}}(t,t) + \sum_{\mathbf{k}'} D_0^{a\beta}(\mathbf{k},\mathbf{k}') C^{\beta\alpha}_{\mathbf{k}',-\mathbf{l}}(t,t) + \sum_{\mathbf{k}'} D_0^{\alpha\beta}(-\mathbf{l},-\mathbf{k}') C^{\beta a}_{-\mathbf{k}',\mathbf{k}}(t,t) \tag{3.17}$$

$$= \sum_{\mathbf{p}} \sum_{\mathbf{q}} \delta(\mathbf{k},\mathbf{p},\mathbf{q}) A^{abc}(\mathbf{k},\mathbf{p},\mathbf{q}) C^{b\alpha}_{-\mathbf{p},-\mathbf{l}}(t,t) h^c_{-\mathbf{q}}$$

$$+ \sum_{\mathbf{p}} \sum_{\mathbf{q}} \delta(\mathbf{k},\mathbf{p},\mathbf{q}) K^{abc}(\mathbf{k},\mathbf{p},\mathbf{q}) [<\zeta^b_{-\mathbf{p}}(t)> C^{c\alpha}_{-\mathbf{q},-\mathbf{l}}(t,t) + C^{b\alpha}_{-\mathbf{p},-\mathbf{l}}(t,t) <\zeta^c_{-\mathbf{q}}(t)>]$$

$$+ \sum_{\mathbf{p}} \sum_{\mathbf{q}} \delta(-\mathbf{l},\mathbf{p},\mathbf{q}) A^{\alpha bc}(-\mathbf{l},\mathbf{p},\mathbf{q}) C^{ba}_{-\mathbf{p},\mathbf{k}}(t,t) h^c_{-\mathbf{q}}$$

$$+ \sum_{\mathbf{p}} \sum_{\mathbf{q}} \delta(-\mathbf{l},\mathbf{p},\mathbf{q}) K^{\alpha bc}(-\mathbf{l},\mathbf{p},\mathbf{q}) [<\zeta^b_{-\mathbf{p}}(t)> C^{ca}_{-\mathbf{q},\mathbf{k}}(t,t) + C^{ba}_{-\mathbf{p},\mathbf{k}}(t,t) <\zeta^c_{-\mathbf{q}}(t)>]$$

$$+ \int_{t_o}^{t} ds \sum_{\mathbf{k}'} N^{a\beta}_{\mathbf{k},-\mathbf{k}'}(t,s) R^{\alpha\beta}_{-\mathbf{l},-\mathbf{k}'}(t,s) + \int_{t_o}^{t} ds \sum_{\mathbf{k}'} N^{\alpha\beta}_{-\mathbf{l},\mathbf{k}'}(t,s) R^{a\beta}_{\mathbf{k},\mathbf{k}'}(t,s)$$

$$- \int_{t_o}^{t} ds \sum_{\mathbf{k}'} \eta^{a\beta}_{\mathbf{k},\mathbf{k}'}(t,s) C^{\alpha\beta}_{-\mathbf{l},\mathbf{k}'}(t,s) - \int_{t_o}^{t} ds \sum_{\mathbf{k}'} \eta^{\alpha\beta}_{-\mathbf{l},-\mathbf{k}'}(t,s) C^{a\beta}_{\mathbf{k},-\mathbf{k}'}(t,s).$$

The system of statistical dynamical equations (3.2) for the mean-field, (3.10) for the two-time covariance, (3.15) for the response function and (3.17) for the single-time covariance constitute the IDIA closure in conjunction with the expressions (3.11) to (3.15) for the nonlinear damping self-energy, nonlinear noise, nonlinear noise self-energy and prescribed random forcing.

## 4. Non-Gaussian Noise and Spurious Vertices

Next we consider the inclusion of non-Gaussian noise. The MSR formalism only allowed for Gaussian initial conditions and did not include random terms although Gaussian random forcing was added in a somewhat ad hoc way. Rose (1974) considered additive random noise and random initial conditions and found that the MSR formalism could then be extended by replacing the vertices $\gamma_i^{2\bullet}$ by $\gamma_i^{2\bullet} + \gamma_i^{2\times}$. Here $i$ can be any positive integer but for consistency with the DIA it runs between 1 and 3; also • is a sequence of $i-1$ values of 1 and × is a sequence of $i-1$ values of 2. Rose (1974) called the vertices $\gamma_i^{2\times}$ 'spurious' vertices. Jensen (1981) using the path integral formalism included general random terms as noted following Eq. (2.1) and discussed in more detail in Appendix A, Sections 9.2 and 9.3. As noted there the 'spurious' vertex $\gamma_3^{222}$ results in an addition to the self-energy.

We shall be primarily interested in the case of homogeneous non-Gaussian white noise (Rose 1974, Jensen 1981), for which the self-energy term in Eq. (A.32) can be written in Fourier space as:

$$Q^{a\alpha}_{\mathbf{k},-\mathbf{k}'}(t,s) = \sum_{\mathbf{p}} \sum_{\mathbf{q}} \sum_{\mathbf{p}'} \sum_{\mathbf{q}'} \delta(\mathbf{k},\mathbf{p},\mathbf{q}) K^{abc}(\mathbf{k},\mathbf{p},\mathbf{q}) \tag{4.1}$$

$$\times \delta(\mathbf{k}',\mathbf{p}',\mathbf{q}') H^{\beta\gamma\alpha}(-\mathbf{p}',-\mathbf{q}',-\mathbf{k}';s,s,s) R^{b\beta}_{-\mathbf{p},-\mathbf{p}'}(t,s) R^{c\gamma}_{-\mathbf{q},-\mathbf{q}'}(t,s)$$

$$\equiv {}^Q\Sigma^{a\alpha}_{\mathbf{k},-\mathbf{k}'}(t,s).$$

Here

$$H^{\beta\gamma\alpha}(-\mathbf{p}',-\mathbf{q}',-\mathbf{k}';s',s'',s) = <\tilde{f}_0^\beta(-\mathbf{p}',s') \tilde{f}_0^\gamma(-\mathbf{q}',s'') \tilde{f}_0^\alpha(-\mathbf{k}',s)> \tag{4.2}$$

and $H \sim \gamma_3^{222}$. Thus the above IDIA equations still hold with Eq. (3.12) replaced by

$$N^{a\alpha}_{\mathbf{k},-\mathbf{k}'}(t,s) = S^{a\alpha}_{\mathbf{k},-\mathbf{k}'}(t,s) + Q^{a\beta}_{\mathbf{k},-\mathbf{k}'}(t,s) + F^{a\alpha}_0(\mathbf{k},-\mathbf{k}',t,s). \tag{4.3}$$

We relate the 'spurious' vertex $H \sim \gamma_3^{222}$ to quantum effects in Section 7. If we are just considering homogeneous white noise forcing then $F_0^{a\alpha}$ drops out of the two-time covariance equation ($F_0^{a\alpha} \to 0$ in Eq. (4.3)) and appears just in the single-time covariance equation as an extra term $F_0^{a\alpha}(\mathbf{k},-\mathbf{k},t,t)\delta_{\mathbf{k},\mathbf{l}}$. For now we shall leave the possibility of general Gaussian noise and the spurious vertex contributed by quantum effects.

## 5. Self-Energy Closure Equations

There are important structural differences between the IDIA closure equations (and the MSR and path integral closure equations) and the QDIA closure equations as noted by Frederiksen (2012b). In the IDIA, and MSR and path integral formalisms, the mean-field equation is formally exact if the single-time covariance were known. The covariance and response function equations are however only approximate to a certain order in perturbation theory or interaction coefficient (coupling constant) with this being second order for the IDIA. When the second-order IDIA covariance is inserted in the mean-field equation it then becomes third order in the interaction coefficient. In contrast in the QDIA closure both the mean-field equation and the diagonal two-point covariance and response function equations are second order in the interaction coefficient or bare vertex function. In the QDIA the transient (eddy-eddy) interaction is expressed in terms of self-energies in covariance and response function equations and, importantly, as well in the mean-field equation; this is not the case for the IDIA mean-field equation. The QDIA closure allows unambiguous identification of the self-energy terms that renormalize the prescribed viscosity and forcing terms in the mean-field equation as well as in the covariance and response function equations.

Next, we summarize the Self-Energy closure equations including contributions from non-Gaussian noise. The Self-Energy has similar complexity of interactions to the IDIA, but the mean-field equation, as well as the two-point equations, are second order in the bare vertex. As in the QDIA closure, all the transient (eddy-eddy) interactions are expressed in terms of self-energies.

*5.1 Statistical closure equations*

The Self-Energy closure equations may be obtained from the IDIA equations by using the first order expression for the covariance in Eq. (B.9) to replace the single-time two-point cumulant $C^{bc}_{-\mathbf{p},-\mathbf{q}}(t,t)$ in the mean-field Equation (3.2) and the two-time cumulant $C^{bc}_{-\mathbf{p},-\mathbf{q}}(t,t')$ in the covariance Equation (3.10). Firstly, using Eq. (B.9) in Eq. (3.2) leads to the expression

$$\sum_{\mathbf{p}}\sum_{\mathbf{q}} \delta(\mathbf{k},\mathbf{p},\mathbf{q}) K^{abc}(\mathbf{k},\mathbf{p},\mathbf{q}) C^{bc}_{-\mathbf{p},-\mathbf{q}}(t,t) \tag{5.1}$$

$$= 2\sum_{\mathbf{p}}\sum_{\mathbf{q}}\sum_{\mathbf{k}'}\sum_{\mathbf{p}'}\sum_{\mathbf{q}'} \delta(\mathbf{k},\mathbf{p},\mathbf{q})\delta(\mathbf{k}',\mathbf{p}',\mathbf{q}') K^{abc}(\mathbf{k},\mathbf{p},\mathbf{q}) \int_{t_o}^{t} ds R^{b\beta}_{-\mathbf{p},-\mathbf{p}'}(t,s) C^{c\gamma}_{-\mathbf{q},\mathbf{q}'}(t,s)$$

$$\times \left[ A^{\beta\gamma\alpha}(-\mathbf{p}',-\mathbf{q}',-\mathbf{k}') h^{\alpha}_{\mathbf{k}'} + 2K^{\beta\gamma\alpha}(-\mathbf{p}',-\mathbf{q}',-\mathbf{k}') <\zeta^{\alpha}_{\mathbf{k}'}(s)> \right]$$

$$= -\int_{t_o}^{t} ds \sum_{\mathbf{k}'} \eta^{a\beta}_{\mathbf{k},\mathbf{k}'}(t,s) <\zeta^{\beta}_{\mathbf{k}'}(s)> + \sum_{\mathbf{k}'} h^{\alpha}_{\mathbf{k}'} \int_{t_o}^{t} ds \chi^{a\alpha}_{\mathbf{k},\mathbf{k}'}(t,s),$$

where the eddy-topographic interaction self-energy

$$\chi^{a\alpha}_{\mathbf{k},\mathbf{k}'}(t,s) = \qquad (5.2)$$

$$2\sum_{\mathbf{p}}\sum_{\mathbf{q}}\sum_{\mathbf{p}'}\sum_{\mathbf{q}'} \delta(\mathbf{k},\mathbf{p},\mathbf{q})\delta(\mathbf{k}',\mathbf{p}',\mathbf{q}')K^{abc}(\mathbf{k},\mathbf{p},\mathbf{q})A^{\beta\gamma\alpha}(-\mathbf{p}',-\mathbf{q}',-\mathbf{k}')R^{b\beta}_{-\mathbf{p},-\mathbf{p}'}(t,s)C^{c\gamma}_{-\mathbf{q},\mathbf{q}'}(t,s)$$

$$\equiv {}^{\chi}\Sigma^{a\alpha}_{\mathbf{k},\mathbf{k}'}(t,s),$$

with the eddy-topographic force is given by

$$f^a_H(\mathbf{k},t) = \sum_{\mathbf{k}'} h^\alpha_{\mathbf{k}'} \int_{t_o}^{t} ds \chi^{a\alpha}_{\mathbf{k},\mathbf{k}'}(t,s) \equiv \sum_{\mathbf{q}} h^c_{-\mathbf{q}} \int_{t_0}^{t} ds \chi^{ac}_{\mathbf{k},-\mathbf{q}}(t,s). \qquad (5.3)$$

The nonlinear damping self-energy $\eta^{a\alpha}_{\mathbf{k},\mathbf{k}'}(t,s)$ is again given in Equation (3.11). Then, with the expression (5.1) the mean-field equation, in the Self-Energy closure, becomes second order in the interaction coefficient and is given by:

$$\frac{\partial^n}{\partial t^n}<\zeta^a_{\mathbf{k}}(t)> +\sum_{\mathbf{k}'} D_0^{a\beta}(\mathbf{k},\mathbf{k}')<\zeta^\beta_{\mathbf{k}'}(t)> = \sum_{\mathbf{p}}\sum_{\mathbf{q}}\delta(\mathbf{k},\mathbf{p},\mathbf{q})A^{abc}(\mathbf{k},\mathbf{p},\mathbf{q})<\zeta^b_{-\mathbf{p}}(t)>h^c_{-\mathbf{q}} \qquad (5.4)$$

$$+\sum_{\mathbf{p}}\sum_{\mathbf{q}}\delta(\mathbf{k},\mathbf{p},\mathbf{q})K^{abc}(\mathbf{k},\mathbf{p},\mathbf{q})<\zeta^b_{-\mathbf{p}}(t)><\zeta^c_{-\mathbf{q}}(t)>$$

$$-\int_{t_o}^{t} ds \sum_{\mathbf{k}'} \eta^{a\beta}_{\mathbf{k},\mathbf{k}'}(t,s)<\zeta^\beta_{\mathbf{k}'}(s)> + \bar{f}^a_0(\mathbf{k},t) + f^a_H(\mathbf{k},t).$$

Here, initial contributions to the off-diagonal covariance matrix have not been shown but can be added as described by O'Kane and Frederiksen (2004) and Frederiksen and O'Kane (2005).

Secondly, the expression for $C^{bc}_{-\mathbf{p},-\mathbf{q}}(t,t')$ in Equation (B.9), is used on the right hand side of Equation (3.10) to derive the two-time cumulant equation for the Self-Energy closure:

$$\frac{\partial^n}{\partial t^n} C^{ab}_{\mathbf{k},-\mathbf{l}}(t,t') + \sum_{\mathbf{k}'} D_0^{a\beta}(\mathbf{k},\mathbf{k}')C^{\beta b}_{\mathbf{k}',-\mathbf{l}}(t,t') \qquad (5.5)$$

$$= \int_{t_o}^{t'} ds \sum_{\mathbf{k}'} \{N^{a\beta}_{\mathbf{k},-\mathbf{k}'}(t,s) + P^{a\beta}_{\mathbf{k},-\mathbf{k}'}(t,s)\} R^{b\beta}_{-\mathbf{l},-\mathbf{k}'}(t',s)$$

$$-\int_{t_o}^{t} ds \sum_{\mathbf{k}'} \{\eta^{a\beta}_{\mathbf{k},\mathbf{k}'}(t,s) + \pi^{a\beta}_{\mathbf{k},\mathbf{k}'}(t,s)\} C^{b\beta}_{-\mathbf{l},\mathbf{k}'}(t',s)$$

for $t > t'$. For $t < t'$ the expression in Eq. (3.8) again applies. Eq. (5.5) is for Gaussian initial conditions and with non-Gaussian initial conditions discussed in Appendix A, Section 9.3. In Equation (5.5),

$$P^{a\alpha}_{\mathbf{k},-\mathbf{k}'}(t,s) = \sum_{\mathbf{p}}\sum_{\mathbf{q}}\sum_{\mathbf{p}'}\sum_{\mathbf{q}'} \delta(\mathbf{k},\mathbf{p},\mathbf{q})\delta(\mathbf{k}',\mathbf{p}',\mathbf{q}')C^{b\beta}_{-\mathbf{p},\mathbf{p}'}(t,s) \qquad (5.6)$$

$$\times [2K^{abc}(\mathbf{k},\mathbf{p},\mathbf{q})<\zeta^c_{-\mathbf{q}}(t)> + A^{abc}(\mathbf{k},\mathbf{p},\mathbf{q})h^c_{-\mathbf{q}}]$$

$$\times [2K^{\alpha\beta\gamma}(-\mathbf{k}',-\mathbf{p}',-\mathbf{q}')<\zeta^\gamma_{\mathbf{q}'}(s)> + A^{\alpha\beta\gamma}(-\mathbf{k}',-\mathbf{p}',-\mathbf{q}')h^\gamma_{\mathbf{q}'}]$$

$$\equiv {}^{P}\Sigma^{a\alpha}_{\mathbf{k},-\mathbf{k}'}(t,s),$$

$$\pi_{\mathbf{k},\mathbf{k}'}^{a\alpha}(t,s) = -\sum_{\mathbf{p}}\sum_{\mathbf{q}} \sum_{\mathbf{p}'}\sum_{\mathbf{q}'} \delta(\mathbf{k},\mathbf{p},\mathbf{q})\delta(\mathbf{k}',\mathbf{p}',\mathbf{q}')R_{-\mathbf{p},-\mathbf{p}'}^{b\beta}(t,s) \qquad (5.7)$$

$$\times [2K^{abc}(\mathbf{k},\mathbf{p},\mathbf{q})<\zeta_{-\mathbf{q}}^{c}(t)> + A^{abc}(\mathbf{k},\mathbf{p},\mathbf{q})h_{-\mathbf{q}}^{c}]$$

$$\times [2K^{\beta\alpha\gamma}(-\mathbf{p}',-\mathbf{k}',-\mathbf{q}')<\zeta_{\mathbf{q}'}^{\gamma}(s)> + A^{\beta\alpha\gamma}(-\mathbf{p}',-\mathbf{k}',-\mathbf{q}')h_{\mathbf{q}'}^{\gamma}]$$

$$\equiv -{}^{\pi}\Sigma_{\mathbf{k},\mathbf{k}'}^{a\alpha}(t,s),$$

are nonlinear noise self-energy and nonlinear damping self-energy terms associated with eddy-mean-field and eddy-topographic interactions. Again, $\eta_{\mathbf{k},\mathbf{k}'}^{a\alpha}(t,s)$ and $S_{\mathbf{k},-\mathbf{k}'}^{a\alpha}(t,s)$ are given in Equations (3.11) and (3.12). Both $S_{\mathbf{k},-\mathbf{k}'}^{a\alpha}(t,s)$ and $P_{\mathbf{k},-\mathbf{k}'}^{a\alpha}(t,s)$ are positive semi-definite in the sense of equation (19) of Bowman et al. (1993). Again Eq. (5.5) is for Gaussian initial conditions and with non-Gaussian initial conditions discussed in Appendix A, Section 9.3.

Thirdly, the equation for the response function in the Self-Energy closure is derived in a similar way by using the first order expression for the response function in Equation (B.12) in the first two terms on the right hand side of Equation (3.15). We find

$$\frac{\partial^n}{\partial t^n} R_{\mathbf{k},\mathbf{l}}^{ab}(t,t') + \sum_{\mathbf{k}'} D_0^{a\beta}(\mathbf{k},\mathbf{k}')R_{\mathbf{k}',\mathbf{l}}^{\beta b}(t,t') = -\int_{t'}^{t} ds \sum_{\mathbf{k}'} \{\eta_{\mathbf{k},\mathbf{k}'}^{a\beta}(t,s) + \pi_{\mathbf{k},\mathbf{k}}^{a\beta}(t,s)\}R_{\mathbf{k}',\mathbf{l}}^{\beta b}(s,t') \qquad (5.8)$$

for $t > t'$ with $R_{\mathbf{k},\mathbf{l}}^{ab}(t,t) = \delta_{\mathbf{k},\mathbf{l}}\delta^{ab}$ and $\delta$ is the Kronecker delta function.

The singe-time cumulant equation for the Self-Energy closure then follows on using Eqs. (3.16) and (5.5):

$$\frac{\partial^n}{\partial t^n} C_{\mathbf{k},-\mathbf{l}}^{ab}(t,t) + \sum_{\mathbf{k}'} D_0^{a\beta}(\mathbf{k},\mathbf{k}')C_{\mathbf{k}',-\mathbf{l}}^{\beta b}(t,t) + \sum_{\mathbf{k}'} D_0^{b\beta}(-\mathbf{l},-\mathbf{k}')C_{-\mathbf{k}',\mathbf{k}}^{\beta a}(t,t) \qquad (5.9)$$

$$= \int_{t_o}^{t} ds \sum_{\mathbf{k}'} \{N_{\mathbf{k},-\mathbf{k}'}^{a\beta}(t,s) + P_{\mathbf{k},-\mathbf{k}'}^{a\beta}(t,s)\}R_{-\mathbf{l},-\mathbf{k}'}^{b\beta}(t,s)$$

$$+ \int_{t_o}^{t} ds \sum_{\mathbf{k}'} \{N_{-\mathbf{l},\mathbf{k}'}^{b\beta}(t,s) + P_{-\mathbf{l},\mathbf{k}'}^{b\beta}(t,s)\}R_{\mathbf{k},\mathbf{k}'}^{a\beta}(t,s)$$

$$- \int_{t_o}^{t} ds \sum_{\mathbf{k}'} \{\eta_{\mathbf{k},\mathbf{k}'}^{a\beta}(t,s) + \pi_{\mathbf{k},\mathbf{k}'}^{a\beta}(t,s)\}C_{-\mathbf{l},\mathbf{k}'}^{b\beta}(t,s)$$

$$- \int_{t_o}^{t} ds \sum_{\mathbf{k}'} \{\eta_{-\mathbf{l},-\mathbf{k}'}^{b\beta}(t,s) + \pi_{-\mathbf{l},-\mathbf{k}'}^{b\beta}(t,s)\}C_{\mathbf{k},-\mathbf{k}'}^{a\beta}(t,s).$$

In summary, the Self-Energy closure consists of Equation (5.4) for the mean-field, (5.5) for the two-time covariance, (5.8) for the response function and (5.9) for the single-time covariance together with the expressions (3.11) to (3.13), (5.2), (5.6) and (5.7) for the self-energy terms and Eq. (3.14) for the random noise. Again in the presence of non-Gaussian noise (3-point function) the Self-Energy equations hold with Eq. (3.12) replaced by Eq. (4.3).

## 6. Inhomogeneous QDIA Closure Equations

The QDIA closure equations were derived directly through perturbation theory and formal renormalization (Frederiksen 1999, 2012a) for the case of equations that are first order in the tendency.

However, it is probably instructive to make the connection between the QDIA and the IDIA closure and MSR and path integral formalisms by obtaining it as a simplification of the Self-Energy closure of Section 5.

*6.1 Statistical closure equations*

To obtain the QDIA closure equations we assume that to lowest order the two-point cumulants and response functions are diagonal in Fourier space. That is

$$R_{\mathbf{k},\mathbf{k'}}^{ab}(t,t') = R_{\mathbf{k}}^{ab}(t,t')\delta_{\mathbf{k},\mathbf{k'}}, \tag{6.1a}$$

$$R_{\mathbf{k}}^{ab}(t,t') \equiv R_{\mathbf{k},\mathbf{k}}^{ab}(t,t'), \tag{6.1b}$$

$$C_{\mathbf{k},\mathbf{k'}}^{ab}(t,t') = C_{\mathbf{k}}^{ab}(t,t')\delta_{\mathbf{k},\mathbf{k'}}, \tag{6.1c}$$

$$C_{\mathbf{k}}^{ab}(t,t') \equiv C_{\mathbf{k},-\mathbf{k}}^{ab}(t,t'). \tag{6.1d}$$

Then the Self-Energy closure mean-field Equation (5.4) reduces to

$$\frac{\partial^n}{\partial t^n} <\zeta_{\mathbf{k}}^a(t)> + D_0^{a\beta}(\mathbf{k}) <\zeta_{\mathbf{k}}^\beta(t)> = \sum_{\mathbf{p}}\sum_{\mathbf{q}} \delta(\mathbf{k},\mathbf{p},\mathbf{q}) K^{abc}(\mathbf{k},\mathbf{p},\mathbf{q}) <\zeta_{-\mathbf{p}}^b(t)><\zeta_{-\mathbf{q}}^c(t)>$$
$$+ \sum_{\mathbf{p}}\sum_{\mathbf{q}} \delta(\mathbf{k},\mathbf{p},\mathbf{q}) A^{abc}(\mathbf{k},\mathbf{p},\mathbf{q}) <\zeta_{-\mathbf{p}}^b(t)> h_{-\mathbf{q}}^c \tag{6.2}$$
$$- \int_{t_o}^{t} ds\, \eta_{\mathbf{k}}^{a\beta}(t,s) <\zeta_{\mathbf{k}}^\beta(s)> + \bar{f}_0^a(\mathbf{k},t) + f_H^a(\mathbf{k},t).$$

Again, initial contributions to the off-diagonal covariance matrix (not shown) may be added as described by O'Kane and Frederiksen (2004) and Frederiksen and O'Kane (2005). In Eq. (6.2) the nonlinear eddy-eddy damping self-energy, eddy-topographic force and eddy-topographic interaction self-energy are given by:

$$\eta_{\mathbf{k}}^{a\alpha}(t,s) = -4\sum_{\mathbf{p}}\sum_{\mathbf{q}} \delta(\mathbf{k},\mathbf{p},\mathbf{q}) K^{abc}(\mathbf{k},\mathbf{p},\mathbf{q}) K^{\beta\gamma\alpha}(-\mathbf{p},-\mathbf{q},-\mathbf{k}) R_{-\mathbf{p}}^{b\beta}(t,s) C_{-\mathbf{q}}^{c\gamma}(t,s)$$
$$\equiv -{}^{\eta}\Sigma_{\mathbf{k}}^{a\alpha}(t,s), \tag{6.3a}$$

$$f_H^a(\mathbf{k},t) = h_{\mathbf{k}}^\alpha \int_{t_o}^{t} ds\, \chi_{\mathbf{k}}^{a\alpha}(t,s), \tag{6.3b}$$

$$\chi_{\mathbf{k}}^{a\alpha}(t,s) = 2\sum_{\mathbf{p}}\sum_{\mathbf{q}} \delta(\mathbf{k},\mathbf{p},\mathbf{q}) K^{abc}(\mathbf{k},\mathbf{p},\mathbf{q}) A^{\beta\gamma\alpha}(-\mathbf{p},-\mathbf{q},-\mathbf{k}) R_{-\mathbf{p}}^{b\beta}(t,s) C_{-\mathbf{q}}^{c\gamma}(t,s)$$
$$\equiv {}^{\chi}\Sigma_{\mathbf{k}}^{a\alpha}(t,s). \tag{6.3c}$$

The two-point Self-Energy closure Equation (5.5) becomes

$$\frac{\partial^n}{\partial t^n} C_{\mathbf{k}}^{ab}(t,t') + D_0^{a\beta}(\mathbf{k}) C_{\mathbf{k}}^{\beta b}(t,t') = \int_{t_o}^{t'} ds\, \{N_{\mathbf{k}}^{a\beta}(t,s) + P_{\mathbf{k}}^{a\beta}(t,s)\} R_{-\mathbf{k}}^{b\beta}(t',s)$$
$$- \int_{t_o}^{t} ds\, \{\eta_{\mathbf{k}}^{a\beta}(t,s) + \pi_{\mathbf{k}}^{a\beta}(t,s)\} C_{-\mathbf{k}}^{b\beta}(t',s) \tag{6.4a}$$

for $t > t'$ while for $t < t'$

$$C_{\mathbf{k}}^{a\alpha}(t,t') = C_{-\mathbf{k}}^{\alpha a}(t',t). \tag{6.4b}$$

Equation (6.4a) is valid for Gaussian initial conditions and can be generalized to non-Gaussian and inhomogeneous initial conditions following the approach of O'Kane and Frederiksen (2004) and Frederiksen and O'Kane (2005). In Equation (6.4a), the nonlinear noise

$$N_{\mathbf{k}}^{a\alpha}(t,s) = S_{\mathbf{k}}^{a\alpha}(t,s) + F_0^{a\alpha}(\mathbf{k},t,s) \tag{6.5}$$

if the specified random forcing is Gaussian while for non-Gaussian random forcing it is given by Eq. (6.13) below. Also, the specified random forcing

$$F_0^{a\alpha}(\mathbf{k},t,s) = <\tilde{f}_0^a(\mathbf{k},t)\tilde{f}_0^\alpha(-\mathbf{k},s)>, \tag{6.6a}$$

and the self-energies are

$$S_{\mathbf{k}}^{a\alpha}(t,s) = 2\sum_{\mathbf{p}}\sum_{\mathbf{q}} \delta(\mathbf{k},\mathbf{p},\mathbf{q}) K^{abc}(\mathbf{k},\mathbf{p},\mathbf{q}) K^{\alpha\beta\gamma}(-\mathbf{k},-\mathbf{p},-\mathbf{q}) C_{-\mathbf{p}}^{b\beta}(t,s) C_{-\mathbf{q}}^{c\gamma}(t,s)$$
$$\equiv {}^S\Sigma_{\mathbf{k}}^{a\alpha}(t,s) \tag{6.6b}$$

$$P_{\mathbf{k}}^{a\alpha}(t,s) = \sum_{\mathbf{p}}\sum_{\mathbf{q}} \delta(\mathbf{k},\mathbf{p},\mathbf{q}) C_{-\mathbf{p}}^{b\beta}(t,s) [2K^{abc}(\mathbf{k},\mathbf{p},\mathbf{q})<\zeta_{-\mathbf{q}}^c(t)> + A^{abc}(\mathbf{k},\mathbf{p},\mathbf{q}) h_{-\mathbf{q}}^c]$$
$$\times [2K^{\alpha\beta\gamma}(-\mathbf{k},-\mathbf{p},-\mathbf{q})<\zeta_{\mathbf{q}}^\gamma(s)> + A^{\alpha\beta\gamma}(-\mathbf{k},-\mathbf{p},-\mathbf{q}) h_{\mathbf{q}}^\gamma] \tag{6.6c}$$
$$\equiv {}^P\Sigma_{\mathbf{k}}^{a\alpha}(t,s),$$

$$\pi_{\mathbf{k}}^{a\alpha}(t,s) = -\sum_{\mathbf{p}}\sum_{\mathbf{q}} \delta(\mathbf{k},\mathbf{p},\mathbf{q}) R_{-\mathbf{p}}^{b\beta}(t,s) [2K^{abc}(\mathbf{k},\mathbf{p},\mathbf{q})<\zeta_{-\mathbf{q}}^c(t)> + A^{abc}(\mathbf{k},\mathbf{p},\mathbf{q}) h_{-\mathbf{q}}^c]$$
$$\times [2K^{\beta\alpha\gamma}(-\mathbf{p},-\mathbf{k},-\mathbf{q})<\zeta_{\mathbf{q}}^\gamma(s)> + A^{\beta\alpha\gamma}(-\mathbf{p},-\mathbf{k},-\mathbf{q}) h_{\mathbf{q}}^\gamma] \tag{6.6d}$$
$$\equiv -{}^\pi\Sigma_{\mathbf{k}}^{a\alpha}(t,s)$$

with $\eta_{\mathbf{k}}^{a\alpha}(t,s)$ is given in Equation (6.3a). Of course $P_{\mathbf{k}}^{a\alpha}(t,s) \equiv {}^P\Sigma_{\mathbf{k}}^{a\alpha}(t,s)$ is also positive definite in the sense of Eq. (19) of Bowman et al. (1993) and could be added to the nonlinear noise term.

Again, the Self-Energy closure response function Equation (5.8) reduces to:

$$\frac{\partial^n}{\partial t^n} R_{\mathbf{k}}^{ab}(t,t') + D_0^{a\beta}(\mathbf{k}) R_{\mathbf{k}}^{\beta b}(t,t') = -\int_{t'}^t ds \{\eta_{\mathbf{k}}^{a\beta}(t,s) + \pi_{\mathbf{k}}^{a\beta}(t,s)\} R_{\mathbf{k}}^{\beta b}(s,t') \tag{6.7}$$

for $t > t'$ with $R_{\mathbf{k}}^{ab}(t,t) = \delta^{ab}$ and $\delta^{ab}$ is the Kronecker delta function. Further, Self-Energy closure Equation (5.9) for the single-time two-point cumulant becomes:

$$\frac{\partial^n}{\partial t^n} C_{\mathbf{k}}^{ab}(t,t) + D_0^{a\beta}(\mathbf{k}) C_{\mathbf{k}}^{\beta b}(t,t) + D_0^{b\beta}(-\mathbf{k}) C_{-\mathbf{k}}^{\beta a}(t,t)$$
$$= \int_{t_o}^t ds\{N_{\mathbf{k}}^{a\beta}(t,s) + P_{\mathbf{k}}^{a\beta}(t,s)\} R_{-\mathbf{k}}^{b\beta}(t,s) + \int_{t_o}^t ds\{N_{-\mathbf{k}}^{b\beta}(t,s) + P_{-\mathbf{k}}^{b\beta}(t,s)\} R_{\mathbf{k}}^{a\beta}(t,s) \tag{6.8}$$
$$-\int_{t_o}^t ds\{\eta_{\mathbf{k}}^{a\beta}(t,s) + \pi_{\mathbf{k}}^{a\beta}(t,s)\} C_{-\mathbf{k}}^{b\beta}(t,s) - \int_{t_o}^t ds\{\eta_{-\mathbf{k}}^{b\beta}(t,s) + \pi_{-\mathbf{k}}^{b\beta}(t,s)\} C_{\mathbf{k}}^{a\beta}(t,s).$$

Finally, the first order expression for the off-diagonal two-time cumulant, in the Self-Energy closure, defined in Equation (B.9) reduces to:

$$C_{\mathbf{k},-\mathbf{l}}^{ab}(t,t') = \int_{t_o}^{t} ds R_{\mathbf{k}}^{a\alpha}(t,s) C_{-\mathbf{l}}^{b\beta}(t',s)[A^{\alpha\beta\gamma}(\mathbf{k},-\mathbf{l},\mathbf{l}-\mathbf{k})h_{(\mathbf{k}-\mathbf{l})}^{\gamma}$$
$$+ 2K^{\alpha\beta\gamma}(\mathbf{k},-\mathbf{l},\mathbf{l}-\mathbf{k}) < \zeta_{(\mathbf{k}-\mathbf{l})}^{\gamma}(s) >]$$
$$+ \int_{t_o}^{t'} ds R_{-\mathbf{l}}^{b\beta}(t',s) C_{\mathbf{k}}^{a\alpha}(t,s)[A^{\beta\alpha\gamma}(-\mathbf{l},\mathbf{k},\mathbf{l}-\mathbf{k})h_{(\mathbf{k}-\mathbf{l})}^{\gamma}$$
$$+ 2K^{\beta\alpha\gamma}(-\mathbf{l},\mathbf{k},\mathbf{l}-\mathbf{k}) < \zeta_{(\mathbf{k}-\mathbf{l})}^{\gamma}(s) >].$$

(6.9)

As well, the first order expression for the off-diagonal response function in Equation (B.12) becomes:

$$R_{\mathbf{k},\mathbf{l}}^{ab}(t,t') = \int_{t'}^{t} ds R_{\mathbf{k}}^{a\alpha}(t,s) R_{\mathbf{l}}^{\beta b}(s,t').$$
$$\left\{ A^{\alpha\beta\gamma}(\mathbf{k},-\mathbf{l},\mathbf{l}-\mathbf{k})h_{(\mathbf{k}-\mathbf{l})}^{\gamma} + 2K^{\alpha\beta\gamma}(\mathbf{k},-\mathbf{l},\mathbf{l}-\mathbf{k}) < \zeta_{(\mathbf{k}-\mathbf{l})}^{\gamma}(s) > \right\}.$$

(6.10)

*6.2 Non-Gaussian noise effects*

For the QDIA closure, with homogeneous non-Gaussian white noise, the non-Gaussian noise self-energy (from Eqs. (4.1) and (A.32)) takes the form:

$$Q_{\mathbf{k}}^{a\alpha}(t,s) = \sum_{\mathbf{p}} \sum_{\mathbf{q}} \delta(\mathbf{k},\mathbf{p},\mathbf{q}) K^{abc}(\mathbf{k},\mathbf{p},\mathbf{q}) H^{\beta\gamma\alpha}(-\mathbf{p},-\mathbf{q},-\mathbf{k};s,s,s) R_{-\mathbf{p}}^{b\beta}(t,s) R_{-\mathbf{q}}^{c\gamma}(t,s) \quad (6.11)$$
$$\equiv {}^Q\Sigma_{\mathbf{k}}^{a\alpha}(t,s).$$

Here, the relationships in Eq. (6.1) have been used and

$$H^{\beta\gamma\alpha}(-\mathbf{p},-\mathbf{q},-\mathbf{k};s',s'',s) = < \tilde{f}_0^{\beta}(-\mathbf{p},s') \tilde{f}_0^{\gamma}(-\mathbf{q},s'') \tilde{f}_0^{\alpha}(-\mathbf{k},s) >. \quad (6.12)$$

With

$$N_{\mathbf{k}}^{a\alpha}(t,s) = S_{\mathbf{k}}^{a\alpha}(t,s) + Q_{\mathbf{k}}^{a\alpha}(t,s) + F_0^{a\alpha}(\mathbf{k},t,s) \quad (6.13)$$

we find that Eqs. (6.4) and (6.8) still apply.

Summing up, the QDIA closure equations consist of the mean-field Equation (6.2), the diagonal two-time cumulant Equation (6.4), the diagonal single-time cumulant Equation (6.8), the diagonal response function Equation (6.7), the off-diagonal cumulant Equation (6.9) and the off-diagonal response function Equation (6.10). The associated self-energies are given in Eqs. (6.3) and (6.6) with the random forcing covariance specified in Eq. (6.6a), the non-Gaussian random forcing in Eqs. (6.11) and (6.12) and the nonlinear noise in Eq. (6.13).

## 7. QDIA Closure for Quantum Fields

Next, we document the details of the QDIA closure equations for a quantum field theory with $g\phi^3$ Lagrangian interaction. Then we present an outline of the QDIA closure for the coupled scalar field $\phi$ and auxiliary field $\chi$ in Equation (2.10a).

## 7.1. QDIA for scalar field with Lagrangian with interaction $g\phi^3$

For the single field in the $g\phi^3$ theory we drop the superscripts and use the equivalences

$$\phi_\mathbf{k}(t) \equiv \zeta_\mathbf{k}(t), \tag{7.1}$$

$$j_\mathbf{k}^\phi(t) \equiv f_0(\mathbf{k},t) \tag{7.2}$$

where $\phi_\mathbf{k}(t)$ is the Fourier transform (Eq. (2.11)) of $\phi(\mathbf{x},t)$. In Eq. (2.8)

$$-\nabla^2 + m^2 \to D_0(\mathbf{k}) = \mathbf{k}^2 + m^2 \tag{7.3}$$

and for the QDIA closure of Section 6,

$$A(\mathbf{k},\mathbf{p},\mathbf{q}) = 0. \tag{7.4}$$

The other vertices $K \sim U_3 = \frac{1}{2}\gamma_3^{211}$ and $H \sim \gamma_3^{222}$ become

$$K(\mathbf{k},\mathbf{p},\mathbf{q}) \to -\tfrac{1}{2}g \tag{7.5}$$

$$H(\mathbf{k},\mathbf{p},\mathbf{q};s,s',s'') \to -\tfrac{1}{4}g\hbar^2. \tag{7.6}$$

Then from Section 6, the QDIA closure equations follow. The mean-field Equation (6.2) becomes

$$\left[\frac{\partial^2}{\partial t^2} + \mathbf{k}^2 + m^2\right]<\phi_\mathbf{k}(t)> = -\tfrac{1}{2}g\sum_\mathbf{p}\sum_\mathbf{q}\delta(\mathbf{k},\mathbf{p},\mathbf{q})<\phi_{-\mathbf{p}}(t)><\phi_{-\mathbf{q}}(t)>$$

$$-\int_{t_o}^t ds\,\eta_\mathbf{k}(t,s)<\phi_\mathbf{k}(s)> + \bar{j}_\mathbf{k}^\phi(t) \tag{7.7}$$

where $\bar{j}_\mathbf{k}^\phi(t) \equiv \bar{f}_0(\mathbf{k},t)$ and the self-energy is given by:

$$\eta_\mathbf{k}(t,s) \equiv -^\eta\Sigma_\mathbf{k}(t,s) = -g^2\sum_\mathbf{p}\sum_\mathbf{q}\delta(\mathbf{k},\mathbf{p},\mathbf{q})R_{-\mathbf{p}}(t,s)C_{-\mathbf{q}}(t,s). \tag{7.8}$$

The QDIA Equation (6.4) for the connected two-point function or cumulant reduces to

$$\left[\frac{\partial^2}{\partial t^2} + \mathbf{k}^2 + m^2\right]C_\mathbf{k}(t,t') = \int_{t_o}^{t'} ds\{S_\mathbf{k}(t,s) + P_\mathbf{k}(t,s) + Q_\mathbf{k}(t,s) + F_0(\mathbf{k},t,s)\}R_{-\mathbf{k}}(t',s)$$

$$-\int_{t_o}^t ds\{\eta_\mathbf{k}(t,s) + \pi_\mathbf{k}(t,s)\}C_{-\mathbf{k}}(t',s) \tag{7.9a}$$

for $t > t'$ while for $t < t'$

$$C_\mathbf{k}(t,t') = C_{-\mathbf{k}}(t',t). \tag{7.9b}$$

Here, the covariance of the random part of the prescribed forcing source

$$F_0(\mathbf{k},t,s) = <\tilde{j}_\mathbf{k}^\phi(t)\tilde{j}_{-\mathbf{k}}^\phi(s)> \equiv <\tilde{f}_0(\mathbf{k},t)\tilde{f}_0(-\mathbf{k},s)>, \tag{7.10a}$$

and the self-energies

$$S_\mathbf{k}(t,s) \equiv {}^S\Sigma_\mathbf{k}(t,s) = \tfrac{1}{2}g^2\sum_\mathbf{p}\sum_\mathbf{q}\delta(\mathbf{k},\mathbf{p},\mathbf{q})C_{-\mathbf{p}}(t,s)C_{-\mathbf{q}}(t,s), \tag{7.10b}$$

$$P_\mathbf{k}(t,s) \equiv {}^P\Sigma_\mathbf{k}(t,s) = g^2\sum_\mathbf{p}\sum_\mathbf{q}\delta(\mathbf{k},\mathbf{p},\mathbf{q})C_{-\mathbf{p}}(t,s)<\phi_{-\mathbf{q}}(t)><\phi_\mathbf{q}(s)>, \tag{7.10c}$$

$$\pi_\mathbf{k}(t,s) \equiv -^\pi\Sigma_\mathbf{k}(t,s) = -g^2\sum_\mathbf{p}\sum_\mathbf{q}\delta(\mathbf{k},\mathbf{p},\mathbf{q})R_{-\mathbf{p}}(t,s)<\phi_{-\mathbf{q}}(t)><\phi_\mathbf{q}(s)>, \tag{7.10d}$$

with $\eta_{\mathbf{k}}(t,s)$ is given in Equation (7.8). Also, from Eq. (6.11) the quantum self-energy is given by

$$Q_{\mathbf{k}}(t,s) \equiv {}^Q\Sigma_{\mathbf{k}}(t,s) = \tfrac{1}{8}g^2\hbar^2 \sum_{\mathbf{p}}\sum_{\mathbf{q}} \delta(\mathbf{k},\mathbf{p},\mathbf{q}) R_{-\mathbf{p}}(t,s) R_{-\mathbf{q}}(t,s). \tag{7.11}$$

The nonlinear noise term becomes

$$N_{\mathbf{k}}(t,s) = S_{\mathbf{k}}(t,s) + Q_{\mathbf{k}}(t,s) + F_0(\mathbf{k},t,s). \tag{7.12}$$

As noted in Section 6, $P_{\mathbf{k}}(t,s) \equiv {}^P\Sigma_{\mathbf{k}}(t,s)$ is also positive definite and could be added to the nonlinear noise term.

Again, the QDIA Equation (6.7) for the retarded propagator or response function reduces to:

$$\left[\frac{\partial^2}{\partial t^2} + \mathbf{k}^2 + m^2\right] R_{\mathbf{k}}(t,t') = -\int_{t'}^{t} ds \{\eta_{\mathbf{k}}(t,s) + \pi_{\mathbf{k}}(t,s)\} R_{\mathbf{k}}(s,t') \tag{7.13}$$

for $t > t'$ with $R_{\mathbf{k}}(t,t) = 1$. Further, Equation (6.8) for the connected single-time two-point function or cumulant becomes:

$$\left[\frac{\partial^2}{\partial t^2} + 2(\mathbf{k}^2 + m^2)\right] C_{\mathbf{k}}(t,t)$$

$$= 2\mathcal{R}e \int_{t_o}^{t} ds\{S_{\mathbf{k}}(t,s) + P_{\mathbf{k}}(t,s) + Q_{\mathbf{k}}(t,s) + F_0(\mathbf{k},t,s)\} R_{-\mathbf{k}}(t,s) \tag{7.14}$$

$$-2\mathcal{R}e \int_{t_o}^{t} ds\{\eta_{\mathbf{k}}(t,s) + \pi_{\mathbf{k}}(t,s)\} C_{-\mathbf{k}}(t,s).$$

Finally, the first order expression for the off-diagonal two-time cumulant in Equation (6.9) reduces to

$$C_{\mathbf{k},-\mathbf{l}}(t,t') = -g\left[\int_{t_o}^{t} ds R_{\mathbf{k}}(t,s) C_{-\mathbf{l}}(t',s) <\phi_{(\mathbf{k}-\mathbf{l})}(s)> + \int_{t_o}^{t'} ds R_{-\mathbf{l}}(t',s) C_{\mathbf{k}}(t,s) <\phi_{(\mathbf{k}-\mathbf{l})}(s)>\right] \tag{7.15}$$

and the first order expression for the off-diagonal propagator in Equation (6.10) becomes

$$R_{\mathbf{k},\mathbf{l}}(t,t') = -g\int_{t'}^{t} ds R_{\mathbf{k}}(t,s) R_{\mathbf{l}}(s,t') <\phi_{(\mathbf{k}-\mathbf{l})}(s)>. \tag{7.16}$$

Of course the classical equations are recovered by setting $Q_{\mathbf{k}}(t,s) \to 0$.

*7.2. QDIA for coupled field-auxiliary field equations associated with $\lambda\phi^4$ Lagrangian interaction*

Next, we consider the dynamical Equations (2.10a) for coupled quadratic interactions between a scalar field $\phi$ and an auxiliary field $\chi$ that can be related to the cubic self-interaction of $\phi$ within a $\lambda\phi^4$ Lagrangian interaction theory (Blagoev et al. 2001). These equations illustrate the more complex structure of the QDIA inhomogeneous closure equations when several fields are coupled. We use the equivalences

$$\phi_{\mathbf{k}}(t) \equiv \zeta_{\mathbf{k}}^1(t); \ \chi_{\mathbf{k}}(t) \equiv \zeta_{\mathbf{k}}^2(t), \tag{7.17}$$

$$j_{\mathbf{k}}^\phi(t) \equiv f_0^1(\mathbf{k},t); \ j_{\mathbf{k}}^\chi(t) \equiv f_0^2(\mathbf{k},t), \tag{7.18}$$

for the Fourier transforms of the variables in Eq. (2.10). Also

$$-\nabla^2 + m^2 \to D_0^{11}(\mathbf{k}) = \mathbf{k}^2 + m^2; -\nabla^2 + M^2 \to D_0^{22}(\mathbf{k}) = \mathbf{k}^2 + M^2, \tag{7.19a}$$

and
$$D_0^{12}(\mathbf{k}) = 0 = D_0^{21}(\mathbf{k}). \tag{7.19b}$$

Again, in the QDIA closure for this system of equations
$$A^{abc}(\mathbf{k},\mathbf{p},\mathbf{q}) = 0 \tag{7.20}$$

and the other vertices $K \sim U_3 = \frac{1}{2}\gamma_3^{211}$ and $H \sim \gamma_3^{222}$ become
$$K^{abc}(\mathbf{k},\mathbf{p},\mathbf{q}) \rightarrow -\tfrac{1}{2}g\delta^{abc}, \tag{7.21}$$
$$H^{abc}(\mathbf{k},\mathbf{p},\mathbf{q};s,s',s'') \rightarrow -\tfrac{1}{4}g\hbar^2\delta^{abc}, \tag{7.22}$$

where
$$\delta^{abc} = \begin{cases} 1 \text{ if } (abc) = (112) \text{ or } (121) \text{ or } (211), \\ \quad 0 \text{ otherwise.} \end{cases} \tag{7.23}$$

With the above substitutions, the QDIA closure of Section 6 then applies for the coupled quadratic interactions between a scalar field $\phi$ and an auxiliary field $\chi$ given in Eq. (2.10a). The direct statistical dynamical analysis of $\lambda\phi^4$ Lagrangian field theories and comparison with the auxiliary field approach, where the bare propagator for the $\chi$ field needs to be replaced (Section 2.2b) prior to renormalization, is to be considered in future work.

## 8. Discussion and Conclusions

We have summarized the derivation of the inhomogeneous IDIA equations for classical field theories including non-Gaussian additive noise and for quantum field theories that have quadratic nonlinearity (cubic Lagrangian) and are first or second order in the tendency. Our approach has used results from the MSR, path integral and Schwinger-Keldysh CTP formulations; in these approaches the mean field equations would be exact if the single-time covariance were known. In the IDIA the covariance and response function equations are second order in the interaction coefficient or coupling constant with the mean field equation then becoming third order in the interaction coefficient. As well, the renormalization of some terms in the IDIA is not as transparent as in the QDIA closure. Here, we explore the relationships between the IDIA and QDIA closure by first modifying the IDIA to form the Self-Energy closure. The Self-Energy closure has a similar structure to the QDIA in that the nonlinear interactions are expressed in terms of self-energy terms and the Self-Energy closure equations are also second order in the interaction coefficient. The QDIA has then been derived by assuming that to lowest order the covariances and response functions are diagonal in Fourier space.

The inhomogeneous QDIA closure equations are much more efficient to compute that the IDIA. In the numerical implementation the resolution or wavenumber (momentum space) cut-off must be specified self-consistently with specified parameters for the viscosity and forcing or the masses and sources. The general subgrid modelling problem has been formulated for the QDIA (Frederiksen 1999, 2012a) and Self-Energy (Frederiksen 2012b) closures. It may also be necessary to regularize the QDIA by using an empirically determined vertex renormalization (Frederiksen and Davies 2004; O'Kane and Frederiksen 2004). For quantum field theories Cooper et al. (2004) also note the need for vertex renormalizations in dimensions greater than 2-space and 1-time (2+1) dimensions.

In applications to classical fluids considerable work has been devoted to establishing relationships between the resolution or cut-off and the required renormalized viscosities and stochastic backscatter (Frederiksen and Davies 1997; Frederiksen and Kepert 2006; O'Kane and Frederiksen 2008; Frederiksen and O'Kane 2008; Zidikheri and Frederiksen 2009, 2010a, b). Indeed, for geophysical fluids of the atmosphere and ocean described by quasi-geostrophic equations, universal scaling laws with cut-off dependence have been established for the renormalized viscosities and backscatter (Kitsios et al. 2012, 2013, 2016; Frederiksen et al. 2017). Again, for quantum fields the relationships between the cut-off and renormalized parameters have been studied in many works (e.g. Cooper et al. 2004; Arrizabalaga et al. 2005; Berges and Wallisch 2017).

In the case of the first order tendency QDIA closure equations the numerical solutions have been made more efficient by using a restart procedure in which the time history integrals are periodically truncated, the three-point cumulant is calculated and employed to specify the non-Gaussian initial conditions in the next integration (Rose 1985; Frederiksen et al. 1994; Frederiksen and Davies 2000; O'Kane and Frederiksen 2004; Frederiksen and O'Kane 2005). In Appendix A, Section 9.3 we have outlined how non-Gaussian initial conditions may be included in the path integral formalism and in the IDIA. However, for the sake of brevity we have not set up a restart procedure here for the second order tendency QDIA closure equations. Interestingly, Juchem et al. (2004) prefer to solve second order tendency statistical equations as a larger system of first order tendency equations in terms of the fields $\phi(1)$ and the canonical momentum fields $\pi(1) = \partial\phi(1)/\partial t_1$. In that case the restart procedure for the QDIA follows exactly as in O'Kane and Frederiksen (2004) and Frederiksen and O'Kane (2005) resulting in the more efficient cumulant update versions of the equations. The QDIA closure offers a computationally tractable means of studying the classical and quantum statistical dynamics of time-dependent non-equilibrium inhomogeneous fields.

## 9. Appendix A: Functional Formalism

The statistical closure equations are formulated using the Schwinger Dyson approach of MSR and the Feynman path integral method of Jensen (1981).

*9.1. Dynamical and statistical dynamical equations*

We consider stochastic differential equations of the form in Eq. (2.1). MSR realized that an equation for the adjoint was also needed for generating the corresponding formalism to the Schwinger-Dyson approach to quantum field theory. This equation is more naturally obtained from the path integral formalism (Jensen 1981). It takes the form

$$(-1)^n \frac{\partial^n}{\partial t_1^n} \hat{\psi}(1) = U_2(2,1)\hat{\psi}(2) + 2U_3(2,3,1)\hat{\psi}(2)\psi(3)$$
$$= U_2(2,1)\hat{\psi}(2) + U_3(2,3,1)\hat{\psi}(2)\psi(3) + U_3(3,2,1)\hat{\psi}(3)\psi(2)$$
(A.1)

and we have generalized to include both cases $n=1$ or $2$. In the case of the second order tendency equation $\hat{\psi}(1)$ is the adjoint of the canonical momentum $\hat{\pi}(1)$ as noted in Section 2.2a. As discussed there, the formalism of Cooper et al. (2001) has the time-derivative and nonlinear term on the same side of the equation, unlike our Eq. (2.1). For our parallel development of the statistical dynamical equations for first and second-order tendency equations it is however more convenient to keep the form (2.1) which changes the sign of the $U_3$ and $\gamma_3$ vertices from those of Cooper et al. (2001) and Blagoev et al. (2001).

Now we define the Pauli matrices

$$\sigma_1 = \begin{pmatrix} 0 & 1 \\ 1 & 0 \end{pmatrix}; \sigma_2 = \begin{pmatrix} 0 & -i \\ i & 0 \end{pmatrix}; \sigma_3 = \begin{pmatrix} 1 & 0 \\ 0 & -1 \end{pmatrix}$$
(A.2)

and also define

$$\rho^{(1)} = -i\sigma_2 = \begin{pmatrix} 0 & -1 \\ 1 & 0 \end{pmatrix}; \rho^{(2)} = \sigma_1 = \begin{pmatrix} 0 & 1 \\ 1 & 0 \end{pmatrix}.$$
(A.3)

Then with

$$\Phi = \begin{pmatrix} \Phi_{s=1} \\ \Phi_{s=2} \end{pmatrix} \equiv \begin{pmatrix} \psi \\ \hat{\psi} \end{pmatrix},$$
(A.4a)

where the subscript is the spinor $s=1$ for $\psi = \Phi_1$ and $s=2$ for $\hat{\psi} = \Phi_2$. Also we note that

$$-i\sigma_2\Phi = \rho^{(1)}\Phi = \begin{pmatrix} -\hat{\psi} \\ \psi \end{pmatrix},$$

$$\sigma_1\Phi = \rho^{(2)}\Phi = \begin{pmatrix} \hat{\psi} \\ \psi \end{pmatrix}.$$

(A.4b)

Now, extend the vectors to include the spinor indices: $\mathbf{1} = (\mathbf{S}_1, \mathbf{x}_1, a_1, t_1) = (\mathbf{S}_1, 1) = (\mathbf{S}_1, \mathbf{1}, t_1)$. Then we rewrite Equations (2.1) and (A.1) as

$$\tau^n(\mathbf{1,2})\Phi(\mathbf{2}) = \gamma_1(\mathbf{1}) + \gamma_2(\mathbf{1,2})\Phi(\mathbf{2}) + \tfrac{1}{2}\gamma_3(\mathbf{1,2,3})\Phi(\mathbf{2})\Phi(\mathbf{3}) \tag{A.5}$$

where

$$\tau^n(\mathbf{1,2}) = \rho^{(n)}(\mathbf{S}_1,\mathbf{S}_2)\delta(\mathbf{x}_1 - \mathbf{x}_2)\delta_{a_1 a_2}\frac{\partial^n}{\partial t_1^n}\delta(t_1 - t_2). \tag{A.6}$$

In terms of the fields $\psi(1)$ and $\hat{\psi}(1)$ we have:

$$\frac{\partial^n}{\partial t_1^n}\psi(1) = \gamma_1^2(1) + \gamma_2^{21}(1,2)\psi(2) + \gamma_2^{22}(1,2)\hat{\psi}(2) + \tfrac{1}{2}\gamma_3^{211}(1,2,3)\psi(2)\psi(3)$$
$$+ \tfrac{1}{2}\gamma_3^{212}(1,2,3)\psi(2)\hat{\psi}(3) + \tfrac{1}{2}\gamma_3^{221}(1,2,3)\hat{\psi}(2)\psi(3) + \tfrac{1}{2}\gamma_3^{222}(1,2,3)\hat{\psi}(2)\hat{\psi}(3) \tag{A.7}$$

and

$$(-1)^n\frac{\partial^n}{\partial t_1^n}\hat{\psi}(1) = \gamma_1^1(1) + \gamma_2^{12}(1,2)\hat{\psi}(2) + \gamma_2^{11}(1,2)\psi(2) + \tfrac{1}{2}\gamma_3^{122}(1,2,3)\hat{\psi}(2)\hat{\psi}(3)$$
$$+ \tfrac{1}{2}\gamma_3^{121}(1,2,3)\hat{\psi}(2)\psi(3) + \tfrac{1}{2}\gamma_3^{112}(1,2,3)\psi(2)\hat{\psi}(3) + \tfrac{1}{2}\gamma_3^{111}(1,2,3)\psi(2)\psi(3) \tag{A.8}$$

Here

$$\gamma_1^2(1) = U_1(1); \gamma_1^1(1) = 0,$$
$$\gamma_2^{21}(1,2) = U_2(1,2); \gamma_2^{22}(1,2) = \text{to be specified for noise effects},$$
$$\gamma_2^{11}(1,2) = U_2(2,1); \gamma_2^{12}(1,2) = 0,$$
$$\gamma_3^{211}(1,2,3) = 2U_3(1,2,3); \gamma_3^{212}(1,2,3) = 0; \gamma_3^{221}(1,2,3) = 0, \tag{A.9}$$
$$\gamma_3^{222}(1,2,3) \text{ to be specified for non - Gaussian noise or quantum effects},$$
$$\gamma_3^{111}(1,2,3) = 0; \ \gamma_3^{112}(1,2,3) = 2U_3(3,1,2),$$
$$\gamma_3^{121}(1,2,3) = 2U_3(2,1,3); \gamma_3^{122}(1,2,3) = 0.$$

Here we note that for $\gamma_1$, $\gamma_2$ and $\gamma_3$ the first superscript refers to the conjugate spinor field while the later superscripts refer to the correct spinor field.

From Equations (A.5) and (A.6), the MSR and path integral formalisms then generate the statistical dynamical equations for

$$G_1(\mathbf{1}) = <\Phi(\mathbf{1})>; G_2(\mathbf{1,2}) = <\Phi(\mathbf{1})\Phi(\mathbf{2})> - G_1(\mathbf{1})G_1(\mathbf{2}). \tag{A.10}$$

These equation take the form

$$\tau^n(\mathbf{1,2})G_1(\mathbf{2}) = \gamma_1(\mathbf{1}) + \gamma_2(\mathbf{1,2})G_1(\mathbf{2}) + \tfrac{1}{2}\gamma_3(\mathbf{1,2,3})[G_2(\mathbf{2,3}) + G_1(\mathbf{2})G_1(\mathbf{3})] \tag{A.11}$$

and

$$\begin{aligned}\tau^n(\mathbf{1,2})G_2(\mathbf{2,1'}) &= \delta(\mathbf{1,1'}) + [\gamma_2(\mathbf{1,2}) + \gamma_3(\mathbf{1,2,3})G_1(\mathbf{3})]G_2(\mathbf{2,1'}) + \tfrac{1}{2}\gamma_3(\mathbf{1,2,3})G_3(\mathbf{3,2,1'}) \\ &= \delta(\mathbf{1,1'}) + [\gamma_2(\mathbf{1,2}) + \gamma_3(\mathbf{1,2,3})G_1(\mathbf{3})]G_2(\mathbf{2,1'}) + \Sigma(\mathbf{1,2})G_2(\mathbf{2,1'}).\end{aligned} \tag{A.12}$$

Here, the self-energy satisfies the relationship

$$\Sigma(\mathbf{1,2})G_2(\mathbf{2,1'}) = \tfrac{1}{2}\gamma_3(\mathbf{1,2,3})G_3(\mathbf{3,2,1'}) \tag{A.13a}$$

or

$$\Sigma(\mathbf{1,1'}) = \tfrac{1}{2}\gamma_3(\mathbf{1,2,3})G_2(\mathbf{2,2'})G_2(\mathbf{3,3'})\Gamma_3(\mathbf{3',2',1'}) \tag{A.13b}$$

where the renormalized vertex function is

$$\Gamma_3(\mathbf{1,2,3}) \equiv -\frac{\delta G_2^{-1}(\mathbf{1,2})}{\delta G_1(\mathbf{3})} = \gamma_3(\mathbf{1,2,3}) + \frac{\delta \Sigma(\mathbf{1,2})}{\delta G_1(\mathbf{3})}. \tag{A.14}$$

The direct interaction approximation (DIA, Kraichnan 1959, 1964, 1972), or bare vertex approximation, corresponds to

$$\Gamma_3^{DIA}(\mathbf{1,2,3}) \approx \gamma_3(\mathbf{1,2,3}) \tag{A.15a}$$

and

$$\Sigma^{DIA}(\mathbf{1,1'}) = \tfrac{1}{2}\gamma_3(\mathbf{1,2,3})G_2(\mathbf{2,2'})G_2(\mathbf{3,3'})\gamma_3(\mathbf{3',2',1'}). \tag{A.15b}$$

Next, we write the DIA equations in component form. The mean field equation is

$$\frac{\partial^n G_1^1(1)}{\partial t_1^n} = U_1(1) + U_2(1,2)G_1^1(2) + U_3(1,2,3)[G_2^1(2,3) + G_1^1(2)G_1^1(3)], \tag{A.16}$$

the covariance equation is

$$\frac{\partial^n G_2^{11}(1,1')}{\partial t_1^n} = [U_2(1,2) + 2U_3(1,2,3)G_1^1(3)]G_2^{11}(2,1') \\ + \Sigma^{21}(1,2)G_2^{11}(2,1') + \Sigma^{22}(1,2)G_2^{12}(1',2), \tag{A.17}$$

and the retarded response function ($t_1 \geq t_1'$) equation is

$$\frac{\partial^n G_2^{12}(1,1')}{\partial t_1^n} = \delta_{a_1 a_2}\delta(\mathbf{x}_1 - \mathbf{x}_1')\delta(t_1 - t_1') + [U_2(1,2) + 2U_3(1,2,3)G_1^1(3)]G_2^{12}(2,1') \\ + \Sigma^{21}(1,2)G_2^{12}(2,1'). \quad (A.18)$$

The expressions for the DIA self-energies are as follows:

$$\Sigma^{21}(1,1') = \gamma_3^{211}(1,2,3)G_2^{12}(2,2')G_2^{11}(3,3')\gamma_3^{121}(3',2',1') \quad (A.19)$$

where the properties (A.9) of the vertex functions have been used and

$$\Sigma^{22}(1,1') = \tfrac{1}{2}\gamma_3^{211}(1,2,3)G_2^{11}(2,2')G_2^{11}(3,3')\gamma_3^{112}(3',2',1') \\ + \tfrac{1}{2}\gamma_3^{211}(1,2,3)G_2^{12}(2,2')G_2^{12}(3,3')\gamma_3^{222}(3',2',1') \\ + \tfrac{1}{2}\gamma_3^{222}(1,2,3)G_2^{21}(2,2')G_2^{21}(3,3')\gamma_3^{112}(3',2',1'). \quad (A.20)$$

Now we suppose that for the problems of interest here

$$U_1(1) = \overline{U}_1(1) + \tilde{U}_1(1),$$
$$U_2(1,2) = \overline{U}_2(\mathbf{1,2})\delta(t_1 - t_2),$$
$$U_3(1,2,3) = \overline{U}_3(\mathbf{1,2,3})\delta(t_1 - t_2)\delta(t_2 - t_3), \quad (A.21)$$
$$\tilde{U}_2(1,2) = 0 = \tilde{U}_3(1,2,3).$$

so that only additive noise $\tilde{U}_1(1)$ and a random contribution to the initial conditions $\tilde{\psi}_0(1)$ are considered.

The contributions to the self-energies from the deterministic components are

$$\Sigma^{21}(1,1') = \overline{\Sigma}^{21}(1,1') = 4\overline{U}_3(\mathbf{1,2,3})G_2^{12}(\mathbf{2},t_1;\mathbf{2'},t_1')G_2^{11}(\mathbf{3},t_1;\mathbf{3'},t_1')\overline{U}_3(\mathbf{3',2',1'}), \quad (A.22)$$

where the random terms do not contribute to $\Sigma^{21}(1,1')$ and

$$\Sigma^{22}(1,1') \approx \overline{\Sigma}^{22}(1,1') = 2\overline{U}_3(\mathbf{1,2,3})G_2^{11}(\mathbf{2},t_1;\mathbf{2'},t_1')G_2^{11}(\mathbf{3},t_1;\mathbf{3'},t_1')\overline{U}_3(\mathbf{1',2',3'}). \quad (A.23)$$

Thus, the mean field equation takes the form

$$\frac{\partial^n G_1^1(1)}{\partial t_1^n} = \overline{U}_1(1) + \overline{U}_2(\mathbf{1,2})G_1^1(\mathbf{2},t_1) + \overline{U}_3(\mathbf{1,2,3})[G_2^1(\mathbf{2},t_1;\mathbf{3},t_1) + G_1^1(\mathbf{2},t_1)G_1^1(\mathbf{3},t_1)], \quad (A.24)$$

the covariance equation becomes

$$\frac{\partial^n G_2^{11}(1,1')}{\partial t_1^n} = [\overline{U}_2(\mathbf{1,2}) + 2\overline{U}_3(\mathbf{1,2,3})G_1^1(\mathbf{3},t_1)]G_2^{11}(\mathbf{2},t_1;1') \\ + \Sigma^{21}(1,2)G_2^{11}(2,1') + \Sigma^{22}(1,2)G_2^{12}(1',2), \quad (A.25)$$

and the (retarded) response function equation is

$$\frac{\partial^n G_2^{12}(1,1')}{\partial t_1^n} = \delta_{a_1 a_2}\delta(\mathbf{x}_1 - \mathbf{x}_1')\delta(t_1 - t_1') + [\overline{U}_2(\mathbf{1,2}) + 2\overline{U}_3(\mathbf{1,2,3})G_1^1(\mathbf{3},t_1)]G_2^{12}(\mathbf{2},t_1;1') \\ + \Sigma^{21}(1,2)G_2^{12}(2,1'). \quad (A.26)$$

*9.2. Non-Gaussian noise*

Next we consider the contribution to the self-energy $\Sigma^{22}(1,1')$ from non-Gaussian noise, and in subsection 9.3, from non-Gaussian initial conditions. Jensen (1981) shows in his Equations (2.49), (2.50) and (3.1) how random noise and random initial conditions add to the Hamiltonian associated with the dynamical equations.

As well the additive random noise adds to the right hand side of the two-point cumulant tendency equation (A.25) the term

$$\frac{\partial^n G_2^{11}(1,1')}{\partial t_1^n} = \text{usual} + F_0(1,2)G_2^{12}(1',2), \qquad (A.27)$$
$$F_0(1,1') = \gamma_2^{22}(1,1') = \ll \tilde{U}_1(1)\tilde{U}_1(1') \gg .$$

Secondly, to the order of the DIA the three-point function of additive random noise contributes to the "spurious" vertex $\gamma^{222}(1,2,3)$ as follows:

$$\gamma_3^{222}(1,2,3) = \ll \tilde{U}_1(1)\tilde{U}_1(2)\tilde{U}_1(3) \gg . \qquad (A.28)$$

Thus

$$^Q\Sigma^{22}(1,1') = {}^{Q_1}\Sigma^{22}(1,1') + {}^{Q_2}\Sigma^{22}(1,1') \qquad (A.29)$$

where

$$\begin{aligned}
{}^{Q_1}\Sigma^{22}(1,1') &= \tfrac{1}{2}\gamma_3^{211}(1,2,3)G_2^{12}(2,2')G_2^{12}(3,3')\gamma_3^{222}(3',2',1') \\
&= \overline{U}_3(\mathbf{1,2,3})G_2^{12}(\mathbf{2},t_1;2')G_2^{12}(\mathbf{3},t_1,3')[\ll \tilde{U}_1(3')\tilde{U}_1(2')\tilde{U}_1(1') \gg]
\end{aligned} \qquad (A.30)$$

and

$$\begin{aligned}
{}^{Q_2}\Sigma^{22}(1,1') &= \tfrac{1}{2}\gamma_3^{222}(1,2,3)G_2^{21}(2,2')G_2^{21}(3,3')\gamma_3^{112}(3',2',1') \\
&= \ll \tilde{U}_1(1)\tilde{U}_1(2)\tilde{U}_1(3) \gg G_2^{21}(2;\mathbf{2}',t_1')G_2^{21}(3;\mathbf{3}',t_1')\overline{U}_3(\mathbf{1',3',2'}).
\end{aligned} \qquad (A.31)$$

For non-Gaussian white noise

$$\begin{aligned}
{}^Q\Sigma^{22}(1,2'')G_2^{12}(1',2'') &= {}^{Q_1}\Sigma^{22}(1,2'')G_2^{12}(1',2'') \\
&= \overline{U}_3(\mathbf{1,2,3})G_2^{12}(\mathbf{2},t_1;2')G_2^{12}(\mathbf{3},t_1,3') \ll \tilde{U}_1(3')\tilde{U}_1(2')\tilde{U}_1(2'') \gg \delta(t_2''-t_2')\delta(t_2''-t_3')G_2^{12}(1',2'') \\
&= \overline{U}_3(\mathbf{1,2,3})G_2^{12}(\mathbf{2},t_1;\mathbf{2'},t_2'')G_2^{12}(\mathbf{3},t_1;\mathbf{3'},t_2'') \ll \tilde{U}_1(3',t_2'')\tilde{U}_1(2',t_2'')\tilde{U}_1(2'') \gg G_2^{12}(1',2'').
\end{aligned} \qquad (A.32)$$

since

$$^{Q_2}\Sigma^{22}(1,2'')G_2^{12}(1',2'')$$
$$=<<\tilde{U}_1(1)\tilde{U}_1(2)\tilde{U}_1(3)>>\delta(t_1-t_2)\delta(t_1-t_3)G_2^{21}(\mathbf{2},t_2;\mathbf{2'},t_2'')G_2^{21}(\mathbf{3},t_3;\mathbf{3'},t_2'')$$
$$\times \overline{U}_3(\mathbf{2'',3',2'})G_2^{12}(1',2'')$$
$$=<<\tilde{U}_1(1)\tilde{U}_1(\mathbf{2},t_1)\tilde{U}_1(\mathbf{3},t_1)>>G_2^{21}(\mathbf{2},t_1;\mathbf{2'},t_2'')G_2^{21}(\mathbf{3},t_1;\mathbf{3'},t_2'')\theta(t_2''-t_1)$$
$$\times \overline{U}_3(\mathbf{2'',3',2'})G_2^{12}(1',2'')\theta(t_1'-t_2'')$$
$$= 0 \text{ because } t_1 \geq t_1'.$$
(A.33)

### 9.3. Non-Gaussian initial conditions

Random initial conditions determine the initial covariance through

$$G_2^{11}(\mathbf{1},t_0;\mathbf{1'},t_0) = <<\tilde{\psi}_0(\mathbf{1},t_0)\tilde{\psi}_0(\mathbf{1'},t_0)>> \tag{A.34}$$

where the double angular bracket denotes the cumulant. Again, to the order of the DIA the three-point function of the (independent) random initial conditions contributes to the "spurious" vertex $\gamma_3^{222}$ as follows:

$$\gamma_3^{222}(1,2,3) = <<\tilde{\psi}_0(1)\tilde{\psi}_0(2)\tilde{\psi}_0(3)>>\delta(t_1-t_0)\delta(t_2-t_0)\delta(t_3-t_0). \tag{A.35}$$

Thus

$$^{I}\Sigma^{22}(1,1') = {}^{I_1}\Sigma^{22}(1,1') + {}^{I_2}\Sigma^{22}(1,1') \tag{A.36}$$

where

$$^{I_1}\Sigma^{22}(1,1') = \tfrac{1}{2}\gamma_3^{211}(1,2,3)G_2^{12}(2,2')G_2^{12}(3,3')\gamma_3^{222}(3',2',1')$$
$$= \overline{U}_3(\mathbf{1,2,3})G_2^{12}(\mathbf{2},t_1;\mathbf{2'})G_2^{12}(\mathbf{3},t_1,\mathbf{3'})<<\tilde{\psi}_0(3')\tilde{\psi}_0(2')\tilde{\psi}_0(1')>>\delta(t_3'-t_0)\delta(t_2'-t_0)\delta(t_1'-t_0)$$
(A.37)

or

$$^{I_1}\Sigma^{22}(1;\mathbf{1'},t_0) = \overline{U}_3(\mathbf{1,2,3})G_2^{12}(\mathbf{2},t_1;\mathbf{2'},t_0)G_2^{12}(\mathbf{3},t_1,\mathbf{3'},t_0)<<\tilde{\psi}_0(\mathbf{3'},t_0)\tilde{\psi}_0(\mathbf{2'},t_0)\tilde{\psi}_0(\mathbf{1'},t_0)>>. \tag{A.38}$$

Note that because of the delta function $\delta(t_1'-t_0)$, $^{I_1}\Sigma^{22}(1,1') = 0$ unless $t_1' = t_0$ or $^{I_1}\Sigma^{22}(1,2'') = 0$ unless $t_2'' = t_0$. Thus,

$$^{I_1}\Sigma^{22}(1,2'')G_2^{12}(1',2'') = {}^{I_1}\Sigma^{22}(1;\mathbf{2''},t_0)G_2^{12}(\mathbf{1'};\mathbf{2''},t_0)$$
$$= \overline{U}_3(\mathbf{1,2,3})G_2^{12}(\mathbf{2},t_1;\mathbf{2'},t_0)G_2^{12}(\mathbf{3},t_1;\mathbf{3'},t_0)<<\tilde{\psi}_0(\mathbf{3'},t_0)\tilde{\psi}_0(\mathbf{2'},t_0)\tilde{\psi}_0(\mathbf{2''},t_0)>>G_2^{12}(\mathbf{1'},t_1';\mathbf{2''},t_0).$$
(A.39)

We also have the second contribution

$$^{I_2}\Sigma^{22}(1,1') = \tfrac{1}{2}\gamma_3^{222}(1,2,3)G_2^{21}(2,2')G_2^{21}(3,3')\gamma_3^{112}(3',2',1')$$
$$= <<\tilde{\psi}_0(1)\tilde{\psi}_0(2)\tilde{\psi}_0(3)>>\delta(t_1-t_0)\delta(t_2-t_0)\delta(t_2-t_0)G_2^{21}(2;\mathbf{2'},t_1')G_2^{21}(3;\mathbf{3'},t_1')\overline{U}_3(\mathbf{1',3',2'})$$
$$= <<\tilde{\psi}_0(\mathbf{1},t_0)\tilde{\psi}_0(\mathbf{2},t_0)\tilde{\psi}_0(\mathbf{3},t_0)>>G_2^{21}(\mathbf{2},t_0;\mathbf{2'},t_1')G_2^{21}(\mathbf{3},t_0;\mathbf{3'},t_1')\overline{U}_3(\mathbf{1',3',2'}).$$
(A.40)

Note that because of the delta function $\delta(t_1-t_0)$, $^{I_2}\Sigma^{22}(1,1') = 0$ unless $t_1 = t_0$. Thus,

$$^{I_2}\Sigma^{22}(1,2'')G_2^{12}(1',2'') = {}^{I_2}\Sigma^{22}(\mathbf{1},t_0;2'')G_2^{12}(1',2'')$$
$$= <<\tilde{\psi}_0(\mathbf{1},t_0)\tilde{\psi}_0(\mathbf{2},t_0)\tilde{\psi}_0(\mathbf{3},t_0)>> G_2^{21}(\mathbf{2},t_0;2',t_1')G_2^{21}(\mathbf{3},t_0;3',t_1')\overline{U}_3(2'',3',2')G_2^{12}(1',t_1';2'',t_0). \quad (A.41)$$

Also we are considering the equation for $G_2^{11}(1,1')$ when $t_1 \geq t_1'$ and this can only occur when $t_1 = t_1' = t_0$ and so this term only contributes to the initial conditions for $G_2^{11}(1,1')$. Thus

$$^{I}\Sigma^{22}(1,2'')G_2^{12}(1',2'') = {}^{I}\Sigma^{22}(1;2'',t_0)G_2^{12}(1';2'',t_0)$$
$$= \overline{U}_3(\mathbf{1,2,3})G_2^{12}(\mathbf{2},t_1;2',t_0)G_2^{12}(\mathbf{3},t_1;3',t_0) <<\tilde{\psi}_0(3',t_0)\tilde{\psi}_0(2',t_0)\tilde{\psi}_0(2'',t_0)>> G_2^{12}(1',t_1';2'',t_0). \quad (A.42)$$

## 10. Appendix B: Perturbation Theory

In this Appendix we derive the first order expressions for the off-diagonal elements of the covariance and response function matrices in terms of the diagonal elements and the mean-field and topography. We then use these results to derive the Self_energy closure equations, presented in Section 5, as modifications to the IDIA closure equations of Section 3. These expressions for the off-diagonal elements are derived through a formal perturbation theory. In this, the interaction coefficients on the right hand side of Equation (3.3) are supposed to be multiplied by a small parameter $\lambda$. The off-diagonal elements are calculated, are formally renormalized and $\lambda$ is restored back to unity.

We suppose that $\tilde{\zeta}_\mathbf{k}^a$ in Equation (3.3) is expanded in the perturbation series

$$\tilde{\zeta}_\mathbf{k}^a = \tilde{\zeta}_\mathbf{k}^{(0)a} + \lambda \tilde{\zeta}_\mathbf{k}^{(1)a} + \ldots \quad (B.1)$$

where $\hat{\zeta}_\mathbf{k}^{(0)a}(t)$ has a multivariate Gaussian distribution. From Equation (3.3) it then follows that to zero order, we have

$$\frac{\partial^n \tilde{\zeta}_\mathbf{k}^{(0)a}(t)}{\partial t^n} + \sum_{\mathbf{k}'} D_0^{\alpha\beta}(\mathbf{k},\mathbf{k}')\tilde{\zeta}_{\mathbf{k}'}^{(0)\beta}(t) = \tilde{f}_0^a(\mathbf{k},t). \quad (B.2)$$

In fact, $D_0^{\alpha\beta}(\mathbf{k},\mathbf{k}') = D_0^{\alpha\beta}(\mathbf{k})$ when $\mathbf{k}' = \mathbf{k}$ and is zero otherwise but we keep the more general form to make the renormalization of the first order covariance and response functions easier to follow.

The first order expression for the fluctuating field is then

$$\frac{\partial^n \tilde{\zeta}_\mathbf{k}^{(1)a}(t)}{\partial t^n} + \sum_{\mathbf{k}'} D_0^{\alpha\beta}(\mathbf{k},\mathbf{k}')\tilde{\zeta}_{\mathbf{k}'}^{(1)\beta}(t) = \sum_\mathbf{p}\sum_\mathbf{q} \delta(\mathbf{k},\mathbf{p},\mathbf{q})A^{abc}(\mathbf{k},\mathbf{p},\mathbf{q})\tilde{\zeta}_{-\mathbf{p}}^{(0)b}h_{-\mathbf{q}}^c \quad (B.3)$$
$$+ \sum_\mathbf{p}\sum_\mathbf{q} \delta(\mathbf{k},\mathbf{p},\mathbf{q})K^{abc}(\mathbf{k},\mathbf{p},\mathbf{q})[<\zeta_{-\mathbf{p}}^b>\tilde{\zeta}_{-\mathbf{q}}^{(0)c} + \tilde{\zeta}_{-\mathbf{p}}^{(0)b}<\zeta_{-\mathbf{q}}^c>$$
$$+ \tilde{\zeta}_{-\mathbf{p}}^{(0)b}\tilde{\zeta}_{-\mathbf{q}}^{(0)c} - <\tilde{\zeta}_{-\mathbf{p}}^{(0)b}\tilde{\zeta}_{-\mathbf{q}}^{(0)c}>].$$

The formal solution to (B.3) is

$$\tilde{\zeta}_\mathbf{k}^{(1)a}(t) = \int_{t_o}^t ds \sum_{\mathbf{k}'} R_{\mathbf{k},\mathbf{k}'}^{(0)a\alpha}(t,s) \left\{ \sum_\mathbf{p}\sum_\mathbf{q} \delta(\mathbf{k}',\mathbf{p},\mathbf{q})A^{\alpha\beta\gamma}(\mathbf{k}',\mathbf{p},\mathbf{q})\tilde{\zeta}_{-\mathbf{p}}^{(0)\beta}(s)h_{-\mathbf{q}}^\gamma \right. \quad (B.4)$$
$$+ \sum_\mathbf{p}\sum_\mathbf{q} \delta(\mathbf{k}',\mathbf{p},\mathbf{q})K^{\alpha\beta\gamma}(\mathbf{k}',\mathbf{p},\mathbf{q})[<\zeta_{-\mathbf{p}}^\beta(s)>\tilde{\zeta}_{-\mathbf{q}}^{(0)\gamma}(s) + \tilde{\zeta}_{-\mathbf{p}}^{(0)\beta}(s)<\zeta_{-\mathbf{q}}^\gamma(s)>$$
$$\left. + \tilde{\zeta}_{-\mathbf{p}}^{(0)\beta}(s)\tilde{\zeta}_{-\mathbf{q}}^{(0)\gamma}(s) - <\tilde{\zeta}_{-\mathbf{p}}^{(0)\beta}(s)\tilde{\zeta}_{-\mathbf{q}}^{(0)\gamma}(s)>] \right\}$$

where $R_{\mathbf{k},\mathbf{k}'}^{(0)\alpha\beta}(t,s)$ is the bare Greens function corresponding to equation (B.2).

The two-time cumulant can then be written in a perturbation series:
$$C^{ab}_{\mathbf{k},-\mathbf{l}}(t,t') = <\tilde{\zeta}^a_{\mathbf{k}}(t)\tilde{\zeta}^b_{-\mathbf{l}}(t')> \quad\quad (B.5)$$
$$= <\tilde{\zeta}^{(0)a}_{\mathbf{k}}(t)\tilde{\zeta}^{(0)b}_{-\mathbf{l}}(t')> + \lambda <\tilde{\zeta}^{(1)a}_{\mathbf{k}}(t)\tilde{\zeta}^{(0)b}_{-\mathbf{l}}(t')> + \lambda <\tilde{\zeta}^{(0)a}_{\mathbf{k}}(t)\tilde{\zeta}^{(1)b}_{-\mathbf{l}}(t')> + \ldots$$

To order zero
$$C^{(0)ab}_{\mathbf{k},-\mathbf{l}}(t,s) = <\tilde{\zeta}^{(0)a}_{\mathbf{k}}(t)\tilde{\zeta}^{(0)b}_{-\mathbf{l}}(t')> \quad\quad (B.6)$$

and to first order in $\lambda$ the off-diagonal or inhomogeneous contribution is
$$C^{(1)ab}_{\mathbf{k},-\mathbf{l}}(t,t') = <\tilde{\zeta}^{(1)a}_{\mathbf{k}}(t)\tilde{\zeta}^{(0)b}_{-\mathbf{l}}(t')> + <\tilde{\zeta}^{(0)a}_{\mathbf{k}}(t)\tilde{\zeta}^{(1)b}_{-\mathbf{l}}(t')> . \quad\quad (B.7)$$

If we substitute Eq. (B.4) into Eq. (B.7) then
$$C^{(1)ab}_{\mathbf{k},-\mathbf{l}}(t,t') = \int_{t_o}^{t} ds \sum_{\mathbf{k}'} R^{(0)a\alpha}_{\mathbf{k},\mathbf{k}'}(t,s)\{\sum_{\mathbf{p}}\sum_{\mathbf{q}}\delta(\mathbf{k}',\mathbf{p},\mathbf{q})A^{\alpha\beta\gamma}(\mathbf{k}',\mathbf{p},\mathbf{q})<\tilde{\zeta}^{(0)\beta}_{-\mathbf{p}}(s)\tilde{\zeta}^{(0)b}_{-\mathbf{l}}(t')>h^{\gamma}_{-\mathbf{q}} \quad\quad (B.8)$$
$$+\sum_{\mathbf{p}}\sum_{\mathbf{q}}\delta(\mathbf{k}',\mathbf{p},\mathbf{q})K^{\alpha\beta\gamma}(\mathbf{k}',\mathbf{p},\mathbf{q})[<\zeta^{\beta}_{-\mathbf{p}}(s)><\tilde{\zeta}^{(0)\gamma}_{-\mathbf{q}}(s)\tilde{\zeta}^{(0)b}_{-\mathbf{l}}(t')>$$
$$+<\tilde{\zeta}^{(0)\beta}_{-\mathbf{p}}(s)\tilde{\zeta}^{(0)b}_{-\mathbf{l}}(t')><\zeta^{\gamma}_{-\mathbf{q}}(s)>]\}$$
$$+\int_{t_o}^{t'} ds \sum_{\mathbf{k}'} R^{(0)b\alpha}_{-\mathbf{l},\mathbf{k}'}(t',s)\{\sum_{\mathbf{p}}\sum_{\mathbf{q}}\delta(\mathbf{k}',\mathbf{p},\mathbf{q})A^{\alpha\beta\gamma}(\mathbf{k}',\mathbf{p},\mathbf{q})<\tilde{\zeta}^{(0)\beta}_{-\mathbf{p}}(s)\tilde{\zeta}^{(0)a}_{\mathbf{k}}(t)>h^{\gamma}_{-\mathbf{q}}$$
$$+\sum_{\mathbf{p}}\sum_{\mathbf{q}}\delta(\mathbf{k}',\mathbf{p},\mathbf{q})K^{\alpha\beta\gamma}(\mathbf{k}',\mathbf{p},\mathbf{q})[<\zeta^{\beta}_{-\mathbf{p}}(s)><\tilde{\zeta}^{(0)\gamma}_{-\mathbf{q}}(s)\tilde{\zeta}^{(0)a}_{\mathbf{k}}(t)>$$
$$+<\tilde{\zeta}^{(0)\beta}_{-\mathbf{p}}(s)\tilde{\zeta}^{(0)a}_{\mathbf{k}}(t)><\zeta^{\gamma}_{-\mathbf{q}}(s)>]\}.$$

We now use the expression for the diagonal covariances in Eq. (B.6) and perform the formal renormalizations $\lambda \to 1$, $R^{(0)ab}_{\mathbf{k},\mathbf{l}} \to R^{ab}_{\mathbf{k},\mathbf{l}}$, $C^{(0)ab}_{\mathbf{k},-\mathbf{l}} \to C^{ab}_{\mathbf{k},-\mathbf{l}}$. Finally, the off-diagonal elements of the two-point cumulant is given by
$$C^{(1)ab}_{\mathbf{k},-\mathbf{l}}(t,t') = \int_{t_o}^{t} ds \sum_{\mathbf{k}'} R^{a\alpha}_{\mathbf{k},\mathbf{k}'}(t,s)\{\sum_{\mathbf{p}'}\sum_{\mathbf{q}'}\delta(\mathbf{k}',\mathbf{p}',\mathbf{q}')[A^{\alpha\beta\gamma}(\mathbf{k}',\mathbf{p}',\mathbf{q}')h^{\gamma}_{-\mathbf{q}'} \quad\quad (B.9)$$
$$+ 2K^{\alpha\beta\gamma}(\mathbf{k}',\mathbf{p}',\mathbf{q}')<\zeta^{\gamma}_{-\mathbf{q}'}(s)>]C^{\beta b}_{-\mathbf{p}',-\mathbf{l}}(s,t')\}$$
$$+\int_{t_o}^{t'} ds \sum_{\mathbf{k}'} R^{b\alpha}_{-\mathbf{l},\mathbf{k}'}(t',s)\{\sum_{\mathbf{p}'}\sum_{\mathbf{q}'}\delta(\mathbf{k}',\mathbf{p}',\mathbf{q}')[A^{\alpha\beta\gamma}(\mathbf{k}',\mathbf{p}',\mathbf{q}')h^{\gamma}_{-\mathbf{q}'}$$
$$+ 2K^{\alpha\beta\gamma}(\mathbf{k}',\mathbf{p}',\mathbf{q}')<\zeta^{\gamma}_{-\mathbf{q}'}(s)>]C^{a\beta}_{\mathbf{k},-\mathbf{p}'}(t,s)\}.$$

We recall that at zero order $C^{(0)ab}_{\mathbf{k},-\mathbf{l}} = \delta_{\mathbf{k},\mathbf{l}}C^{(0)ab}_{\mathbf{k},-\mathbf{k}}$ is in fact diagonal in spectral space.

The derivation of the expression for the off-diagonal response function proceeds in a similar way. It is defined by
$$R^{ab}_{\mathbf{k},\mathbf{l}}(t,t') = \left\langle \frac{\delta\tilde{\zeta}^a_{\mathbf{k}}(t)}{\delta\tilde{f}^b_{\mathbf{l}}(t')} \right\rangle. \quad\quad (B.10a)$$

and can be written in the perturbation series:
$$R^{ab}_{\mathbf{k},\mathbf{l}}(t,t') = \left\langle \frac{\delta\tilde{\zeta}^{(0)a}_{\mathbf{k}}(t)}{\delta\tilde{f}^b_{\mathbf{l}}(t')} \right\rangle + \lambda\left\langle \frac{\delta\tilde{\zeta}^{(1)a}_{\mathbf{k}}(t)}{\delta\tilde{f}^b_{\mathbf{l}}(t')} \right\rangle + \ldots \quad\quad (B.10b)$$

To order zero

$$R_{\mathbf{k},\mathbf{l}}^{(0)ab}(t,t') = \left\langle \frac{\delta \widetilde{\zeta}_{\mathbf{k}}^{(0)a}(t)}{\delta \widetilde{f}_{\mathbf{l}}^{b}(t')} \right\rangle \tag{B.10c}$$

and to first order in $\lambda$, the general off-diagonal or inhomogeneous contribution is

$$R_{\mathbf{k},\mathbf{l}}^{(1)ab}(t,t') = \int_{t'}^{t} ds \sum_{\mathbf{k}'} R_{\mathbf{k},\mathbf{k}'}^{(0)a\alpha}(t,s) \cdot \left\{ \sum_{\mathbf{p}} \sum_{\mathbf{q}} \delta(\mathbf{k}',\mathbf{p},\mathbf{q}) A^{\alpha\beta\gamma}(\mathbf{k}',\mathbf{p},\mathbf{q}) \left\langle \frac{\delta \widetilde{\zeta}_{-\mathbf{p}}^{(0)\beta}(s)}{\delta \widetilde{f}_{\mathbf{l}}^{b}(t')} \right\rangle h_{-\mathbf{q}}^{\gamma} \right. \tag{B.11}$$

$$\left. + \sum_{\mathbf{p}} \sum_{\mathbf{q}} \delta(\mathbf{k}',\mathbf{p},\mathbf{q}) K^{\alpha\beta\gamma}(\mathbf{k}',\mathbf{p},\mathbf{q}) [<\zeta_{-\mathbf{p}}^{\beta}(s)> \left\langle \frac{\delta \widetilde{\zeta}_{-\mathbf{q}}^{(0)\gamma}(s)}{\delta \widetilde{f}_{\mathbf{l}}^{b}(t')} \right\rangle + \left\langle \frac{\delta \widetilde{\zeta}_{-\mathbf{p}}^{(0)\beta}(s)}{\delta \widetilde{f}_{\mathbf{l}}^{b}(t')} \right\rangle <\zeta_{-\mathbf{q}}^{\gamma}(s)>] \right\}.$$

Again, with the formal renormalizations $\lambda \to 1$, $R_{\mathbf{k},\mathbf{l}}^{(0)ab} \to R_{\mathbf{k},\mathbf{l}}^{ab}$ we find that

$$R_{\mathbf{k},\mathbf{l}}^{(1)ab}(t,t') = \int_{t'}^{t} ds \sum_{\mathbf{k}'} R_{\mathbf{k},\mathbf{k}'}^{a\alpha}(t,s) \{ \sum_{\mathbf{p}'} \sum_{\mathbf{q}'} \delta(\mathbf{k}',\mathbf{p}',\mathbf{q}') [A^{\alpha\beta\gamma}(\mathbf{k}',\mathbf{p}',\mathbf{q}') h_{-\mathbf{q}'}^{\gamma} \tag{B.12}$$

$$+ 2K^{\alpha\beta\gamma}(\mathbf{k}',\mathbf{p}',\mathbf{q}') <\zeta_{-\mathbf{q}'}^{\gamma}(s)>] \} R_{-\mathbf{p}',\mathbf{l}}^{\beta b}(s,t').$$

Note that at zero order $R_{\mathbf{k},\mathbf{l}}^{(0)ab} = \delta_{\mathbf{k},\mathbf{l}} R_{\mathbf{k},\mathbf{k}}^{(0)ab}$ is diagonal in spectral space.